\documentstyle[preprint,aps]{revtex}
\begin{document}
\draft
\preprint{NUSc/93-04}
\title{
Van Hove Exciton-Cageons and High-T$_c$ Superconductivity: XA: \\
Role of Spin-Orbit Coupling in Generating a Diabolical Point}

\author{R.S. Markiewicz}

\address{Physics Department and Barnett Institute,
Northeastern U.,
Boston MA 02115}
\maketitle

\begin{abstract}
Spin-orbit coupling plays a large role in stabilizing the low-temperature
orthorhombic phase of La$_{2-x}$Sr$_x$CuO$_4$.  It splits the degeneracy of
the van Hove singularities (thereby stabilizing the distorted phase) and
completely changes the shape of the Fermi surfaces, potentially introducing
diabolical points into the band structure.  The present paper gives a
detailed account of the resulting electronic structure.
\par
A slave boson calculation shows how these results are modified in the presence
of strong correlation effects.  A scaling regime, found very close to the
metal-insulator transition, allows an analytical determination of the
crossover, in the limit of zero oxygen-oxygen hopping, $t_{OO}\rightarrow 0$.
Extreme care must exercised in chosing the parameters of the
three-band model.  In particular, $t_{OO}$ is renormalized from its LDA
value.  Furthermore, it is suggested that the slave boson model be
spin-corrected, in which case the system is close to a metal-insulator
transition at half filling.

\end{abstract}

\pacs{PACS numbers~:~~71.27.+a, ~74.20.Mn  }

\section{Introduction}

The suggestion that the van Hove singularity (vHs) plays an important role in
high-T$_c$ superconductivity\cite{vHs} has been greatly strengthened by recent
photoemission studies, which find that the Fermi level is very close to the vHs
at the compositions of optimum T$_c$, both in Bi$_2$Sr$_2$CaCu$_2$O$_8$
(Bi-2212)\cite{PE1} and in YBa$_2$Cu$_3$O$_7$ (YBCO)\cite{PE2}.
Furthermore, it has been shown theoretically that the vHs's in
La$_{2-x}$Sr$_x$CuO$_4$ (LSCO) and related compounds drive a very strong
electron-phonon interaction\cite{RM8A,RM8,RM8C}.  The mechanism is via a
vHs-Jahn-Teller effect, wherein the two vHs's provide the electronic degeneracy
which can be split by a lattice distortion -- a band Jahn-Teller (JT) effect.
However, in order to reproduce the experimental phase diagram, it was
found\cite{RM8A} that electronic correlations play an essential role, by
renormalizing the band structure.  The present paper is a first attempt to
treat both the electron-hole coupling and correlation effects on an equal
footing.
\par
It is found that spin-orbit coupling plays a dominant role in driving the
structural instability, and leads to major reconstruction of the Fermi
surfaces.
It acts like a form of umklapp scattering, splitting the degeneracy of the
vHs's
(even in the LTO phase!), and producing gaps in the vicinity of the old vHs's,
thereby transforming the large Fermi surfaces into small pockets in better
agreement with Hall effect measurements.  However, while this coupling
can restructure the Fermi surfaces, {\it it cannot eliminate vHs's}.  Instead,
the vHs's are split and shifted in energy, forming a very complex pattern of
up to four vHs's, with corresponding peaks in the density of states (dos).  A
detailed description of the various possibilities is presented.
\par
A most interesting possibility is that, if spin-orbit coupling is strong
enough, the band may posess a {\it diabolical point}, at which the Fermi
surface shrinks to a point, and the dos vanishes.  The possible role of such a
state in the antiferromagnetic insulating phase at half filling is discussed.
\par
The importance of these features depends critically on relative values
of several band parameters -- both bare parameter values and how the values
renormalize in the presence of correlation effects.  It is found that there
remains considerable uncertainty in the optimal values for several parameters,
but approximate estimates are given.
\par
Finally, correlation effects are included by means of a slave boson
calculation.  In the absence of spin-orbit (and other electron-phonon)
coupling, a full calculation is given (to lowest order in $1/N$) for the
three-band model of the CuO$_2$
planes, and the metal-insulator phase boundary is found.  In the limit of
strong renormalization of the Cu-O hopping parameter, a scaling regime is
found in which the self-consistent equations greatly simplify.  The phase
boundary is found to lie in the range of the optimum parameters.
\par
When spin-orbit and electron-phonon coupling are included, it is found that
there is still a scaling regime, and that, since the additional coupling is
small, the phase boundary is nearly unchanged.  For the parameters chosen, the
state at half filling is semimetallic; however, a larger, renormalized value of
the spin-orbit coupling parameter might be more appropriate.  In this case, the
state at half filling would become a zero-gap semiconductor with diabolical
point.
\par
This paper naturally divides into two parts.  First, Sections II, III show
how spin-orbit (and electron-phonon) coupling modifies the three-band,
tight-binding model of the CuO$_2$ planes, splitting the degeneracy of the
vHs's
{\it even in the LTO phase}.  The new Fermi surfaces and vHs's are described in
detail, and it is shown that at half filling there can be either compensated,
semimetallic Fermi surfaces, or a zero-gap semiconductor with a {\it diabolical
point} at precisely half filling (Section III).  The remainder of the paper
introduces correlations into the problem, via slave boson calculations.  This
includes a discussion of the proper choice of parameters (Section IV), the
self-consistent equations and their modification in the presence of spin-orbit
coupling (Section VA), the analysis of a {\it scaling regime} (and its
modification in the presence of spin-orbit coupling) which exists near the
metal-insulator transition (Section VB,C), and a more general discussion of
parameter renormalization near this transition (Section VD).  Conclusions
are discussed in Section VI, while Section VII discusses a number of new
features which arise in the presence of spin-orbit coupling and the possible
diabolical point.

\section{Role of Spin-Orbit Coupling in the Structural
Transitions}

Spin-orbit coupling plays two major roles in the structural transitions.
First, since spin-orbit coupling is linear in the tilt angle $\theta$, while
ordinary electron-phonon coupling is quadratic, the spin-orbit term
dominates in driving the structural transition\cite{Coff,Bones}.
Secondly, spin-orbit coupling can split the degeneracy of the vHs's, leading
to a large electronic contribution to the stabilization of the LTO
phase\cite{RM8C}.

\subsection{Spin-Orbit Coupling}

In La$_{2-x}$Ba$_x$CuO$_4$ (LBCO), there is a series of structural transitions
as the temperature is lowered, from a high-temperature tetragonal (HTT) phase
near room temperature, to the low-temperature orthorhombic (LTO) phase, and, in
a restricted doping range, to the low-temperature tetragonal phase (LTT), which
has a deleterious effect on superconductivity.  It has long been recognized
that
the distortion in the LTT phase splits the degeneracy of the two
vHs's\cite{LTT}, thereby
acting like a classical charge density wave, or static JT effect.
\par
On the other hand, it was generally assumed\cite{Poug} that since the two vHs's
remain degenerate by symmetry in the LTO phase, the LTO transition is unrelated
to the vHs's.  This argument is incorrect, and in fact there are
two distinct ways in which splitting of the vHs's can drive the LTO
transition\cite{RM8C}.  From general symmetry arguments, it can be shown that
spin-orbit coupling can lift the orbital degeneracy on most of the orthorhombic
Brillouin zone, leading to a splitting of the vHs density of states peak.
Alternatively, the LTO phase could be a {\it dynamic} JT phase, in which the
local symmetry differs from the global one, stabilized by a dynamic splitting
of
the vHs degeneracy.  Based on this picture, I presented\cite{RM8A} a simplified
calculation of the resulting doping dependence of the LTO phase transition and
the competing superconductivity.
\par
A complication in analyzing this transition is that the soft mode of the
transition involves a tilting of the CuO$_6$ octahedra, which has
conventionally
been found to couple only quadratically to the electrons (non-Migdal
behavior)\cite{BB,RM8,SongAn}. Recently, however, it has been pointed out that
spin-orbit coupling also introduces a linear coupling between the electrons and
the tilt mode\cite{Bones}.  In the present paper, I explore the role of
spin-orbit coupling in driving the LTO transition.  This Subection describes
the
modifications of the band structure due to spin-orbit coupling in the LTO and
LTT phases.  Unlike earlier work\cite{Bones}, no magnetic ordering is assumed.
\par
The analysis is based on the standard three-band model of the CuO$_2$ plane,
involving both Cu-O ($t_{CuO}$) and O-O ($t_{OO}$) hopping:
$$H=\sum_j\bigl(\Delta d^{\dagger}_jd_j
+\sum_{\hat\delta}t_{CuO}[d^{\dagger}_jp_{j+\hat\delta}+(c.c.)]
+\sum_{\hat\delta^{\prime}}t_{OO}[p^{\dagger}_{j+\hat\delta}p_{j+\hat\delta^
{\prime}}+(c.c.)]$$
$$+Un_{j\uparrow}n_{j\downarrow}\bigr),\eqno(1)$$
where $d^{\dagger}$ ($p^{\dagger}$) is a creation operator for holes on Cu (O),
j is summed over lattice sites, $\hat\delta$ over nearest neighbors,
$\hat\delta^{\prime}$ over next-nearest (O-O) neighbors, and c.c.
stands for complex conjugate.  Energies are measured from the center of the
O bands, with $\Delta$ the bare Cu-O splitting, and $U$ is the on-site Coulomb
repulsion.  In a slave boson calculation, $U$ produces correlations which
renormalize the one-electron band parameters.  Hence, in the present section
the Coulomb term will be neglected, in order to study how the band structure is
modified as extra terms are added to Eq. 1.  In Section IV, the slave boson
technique will be used to determine the renormalized band structure.
\par
Coffey, et al.\cite{Coff} showed that, in the presence of octahedral tilt,
spin-orbit coupling introduces an additional, spin-dependent hopping term into
the Hamiltonian, Eq. 1, of the form
$$H^{\prime}=\sum_{j,\hat\delta}\sum_{\alpha ,\beta}\bigl(
d^{\dagger}_{j,\alpha}i\vec\lambda_{j,j+\hat\delta}\cdot\vec\sigma_{\alpha ,
\beta}p_{j+\hat\delta ,\beta}+(c.c.)\bigr),\eqno(2)$$
where $\alpha$ and $\beta$ are spin indices, $\vec\sigma_{\alpha ,\beta}$ is
the
vector of spin matrices, and the coupling parameter has the form
$$\vec\lambda_{j,j+\hat x}=(-1)^{j_x+j_y}(\lambda_1,\lambda_2^{\prime},0),
\eqno(3a)$$
$$\vec\lambda_{j,j+\hat y}=-(-1)^{j_x+j_y}(\lambda_2,\lambda_1^{\prime},0),
\eqno(3b)$$
where the $\lambda_i$'s are microscopic parameters, estimated to be $\lambda_i
=\gamma_i\theta$, with $\theta$ the octahedral tilt angle and $\gamma_2
>\gamma_1>0$, $\lambda_i^{\prime} =\lambda_i$ in the LTO phase, and $=0$ in
the LTT phase, and the values of $\lambda_i$ are larger by a factor $\sqrt{2}$
in the LTT phase, for a given tilt angle.

\subsection{Modified Band Structure in the Presence of Spin-Orbit Scattering}

The LTO phase of LSCO has symmetry group Bmab; in the absence of spin-orbit
coupling, this means that the electronic states on the $X^*$ face of the
Brillouin zone have an additional two-fold degeneracy (beyond the usual spin
degeneracy), which means that the degeneracy of
the two vHs's cannot be split by a transition to a uniform orthorhombic phase.
\par
Spin-orbit scattering dramatically alters this situation.  The spin ceases to
be
a good quantum number, and the eigenfunctions mix up and down spin states.
There
is still a two-fold Kramers degeneracy, but the extra degeneracy on the $X^*$
face is lifted everywhere, except at $X^*$ itself\cite{RM8C}.  This section
will
provide a detailed calculation of how the Fermi surfaces are changed by the
spin-orbit coupling, Eq. 2.
\par
The orthorhombic unit cell is double the area of the HTT cell, containing
2 Cu's and 4 O's per layer.  Since up and down spins must be treated
separately,
the resulting Hamiltonian matrix is a $12\times 12$.  By properly symmetrizing
the two Cu's (and corresponding O's) this can be reduced to two $6\times 6$
matrices, which are Kramers' doublets.  The eigenvalues are then given by
$$det\left(\matrix{\Delta_- -E&2tc_x&2tc_y&0&-2\lambda_x^*s_x&-2\lambda_y
^*s_y\cr
   2tc_x&-E&v&2\lambda_x^*c_x&0&0\cr
   2tc_y&v&-E&2\lambda_y^*c_y&0&0\cr
   0&2\lambda_xc_x&2\lambda_yc_y&\Delta_+-E&2ts_x&2ts_y\cr
   -2\lambda_xs_x&0&0&2ts_x&-E&u\cr
   -2\lambda_ys_y&0&0&2ts_y&u&-E}\right)=0,\eqno(4)$$
with $t=t_{CuO}$, $s_i=sin(k_ia/2)$, $c_i=cos(k_ia/2)$, $i=x,y$,
$u=4t_{OO}s_xs_y$, $v=4t_{OO}c_xc_y$, $\Delta_{\pm}=\Delta$,
and, for LTO, $\lambda_x=(\lambda_1+i\lambda_2)$, $\lambda_y=-(\lambda_2+i
\lambda_1)$ (for LTT, $\lambda_x=\sqrt{2}\lambda_1$, $\lambda_y=-\sqrt{2}
\lambda_2$).  For the general case, Eq. 4 must be solved numerically, but in
the special case $t_{OO}=0$, the determinant can be reduced to a quadratic
equation
$$[E(\Delta -E)+4t^2+2\bar\lambda^2]^2=4[t^2(\bar c_x+\bar c_y)+2|\lambda_x|^2
\bar c_x+2|\lambda_y|^2\bar c_y]^2$$
$$-16t^2\bar c_x\bar c_y\lambda^2,\eqno(5)$$
with $\bar\lambda^2=\lambda_1^2+\lambda_2^2$, $\lambda=\lambda_1+\lambda_2$,
and $\bar c_i =cos(k_ia)$, $i=x,y$.
When $\bar c_x=\bar c_y=0$, the right-hand side of Eq. 5 vanishes, leading to
the (symmetry-required) two-fold degeneracy at $X^*$ and $Y^*$ ($\bar M$, in
the
pseudo-tetragonal cell we are using).  The generalization of Eq. 5 to arbitrary
$t_{OO}$ is given in Appendix I.

\subsection{Electron-Phonon Interaction}

As discussed elsewhere\cite{RM8}, ordinary electron-phonon coupling can split
the vHs degeneracy in the LTT phase, but not in a uniform LTO phase.  Moreover,
the coupling is quadratic in the tilt angle, $\theta$, and hence of non-Migdal
form\cite{BB,RM8,SongAn}.  However, after the spin-orbit coupling leads to a
finite tilt, this ordinary
electron-phonon coupling can enhance the tilt instability.  Hence, in this
subsection, the above expressions are modified to include a JT-like phonon
coupling.  In Ref. \cite{RM8}, I showed that an important contribution comes
from coupling between the Cu $d_{x^2-y^2}$ orbital and the O $p_z$ orbital.
This contribution can be incorporated into the interaction matrix, Eq. 4, by
the
substitution (in the LTO phase)
$$\Delta_{\pm}=\Delta +\eta [2\pm (\bar c_x+\bar c_y)],\eqno(6a)$$
where
$$\eta={2t^{\prime 2}\over E},\eqno(6b)$$
and $t^{\prime}$ $(\propto sin\theta )$ is the Cu-O$p_z$ hopping matrix
element.
Note that strictly speaking, $\eta$ is inversely proportional to the energy,
and hence will modify the eigenvalues of Eq. 4 at each angle.  However, it
should be sufficiently accurate for present purposes to treat $\eta$ as a
fixed parameter, of approximate magnitude $\eta\sim 2t^{\prime 2}/\Delta$.  In
this approximation, Equation 6 is very similar to the electron-phonon coupling
recently proposed by Song and Annett\cite{SongAn}.
\par
{}From the form of Eq. 6, the electron-phonon coupling does not split the
vHs degeneracy in the LTO phase -- in fact, Eq. 6 reduces to a constant along
the Brillouin zone diagonal, $\bar c_x+\bar c_y=0$.  In the LTT phase, the
expression should be modified, as
$$2-(\bar c_x+\bar c_y)\rightarrow 2(sin^2\theta_x s_x^2+sin^2\theta_ys_y^2),$$
where $\theta_i$ is the tilt in the LTT phase (e.g., in one domain, $\theta_y$
would vanish).

\subsection{Numerical Results}

\par
Equation 4 has been solved numerically, and the resulting energy bands and
Fermi
surfaces are illustrated in Figures 1-6.  Figure 1 shows the resulting energy
dispersions for a series of different band parameters.  The appropriate choice
of parameters for calculating the effects of correlations will be discussed in
detail in Section IV.  Here, for illustrative purposes, values close to the
bare
band parameters will be assumed.  These can be taken as $t=1.3eV$,
$\Delta =4eV$, $t_{OO}=0.65eV$, $\lambda_2\le\lambda_1\simeq 6meV$, and $\eta
\simeq 5-10meV$.  The special points of the Brillouin zone
are based on the tetragonal supercell, which has twice the area of the
orthorhombic zone: $\Gamma =(0,0)$, $X=(\pi /a,0)$, $Y=(0,\pi /a)$,
$M=(\pi /a,\pi /a)$, and $\bar M =M/2$.
\par
The original, untilted case, with $\lambda_i=\eta =0$, is illustrated in Figure
1a.  It can easily be understood as the usual dispersion found
in the tetragonal cell (solid line) modified by the zone folding associated
with
the real space cell size doubling, which maps $M\rightarrow\Gamma$.  The
corresponding Fermi surfaces are shown in Fig. 2a.  The energy levels freely
intersect without interacting along the orthorhombic zone boundary.  In
particular, they need not be perpendicular to this boundary when they intersect
it.  Under these circumstances, it is convenient to work in a double cell
(the original tetragonal zone) and ignore the zone folded bands.  This is what
is usually done.
\par
However, spin-orbit coupling breaks this degeneracy almost everywhere
along the zone boundary.  This coupling acts like an umklapp term, mixing the
original and zone folded bands.  The resulting, reconstructed Fermi surfaces
bear little resemblance to the original surfaces in the regions of overlap.
Figure 2b illustrates how the Fermi surfaces of Fig. 2a are modified by the
presence of a small spin-orbit term ($\lambda_2=6meV$, $\lambda_1=\eta =0$).
The two lower energy levels ($E_F=4.5,$ 5eV) did not overlap, and are
essentially unchanged.  The third level, 5.28eV, was initially at the vHs, with
area proportional to $1+x$ (with $x$ the Sr doping), but the overlap has
distorted the Fermi surface into a pair of closed hole Fermi surfaces (recall
that the figure illustrates only one-quarter of the tetragonal cell), each of
area $x/2$.  For the fourth energy level, 5.5eV, the reconstruction converts
the original Fermi surface into an electron-like Fermi surface, centered at $X$
and $Y$, and a hole-like surface, centered at $\bar M$ (and symmetry-equivalent
points).  The interaction has split the vHs degeneracy, as can be seen in Fig.
1b, the energy dispersion corresponding to the Fermi surfaces of Fig. 2b.  A
gap
has been opened up at the $X$-point of the zone, the location of the original
vHs.  However, the interactions cannot eliminate a vHs, but only split it into
multiple vHs's.  This will be further discussed in the following subsection.
\par
For convenience in discussing the effects of various parameters on the band
structure, I have used large values of the parameters to enhance the effects
(as will be discussed later, such larger values may indeed be more
appropriate).
Figure 1c shows that when $\lambda_2$ is increased to $50meV$, the
band structure is qualitatively unchanged, with only an enhanced band gap.
As shown in Fig. 1d, the effect of $\lambda_1$ on the gap is roughly additive
with $\lambda_2$, while $\eta$ (dashed lines in Fig. 1c) tends to narrow the
antibonding band and shift it upward (as found earlier\cite{RM8A}) without
modifying the region near the vHs's.  (To avoid an overall band shift, I have
assumed $\Delta +2\eta =4eV$ for the dashed lines.)
\par
All of the above refer to the LTO phase.  The effect of
the $\lambda_i$'s on the LTT phase is nearly identical, as illustrated in Fig.
3a. The electron-phonon parameter $\eta$ has a much larger effect in the LTT
phase\cite{RM8A}, Fig. 3b.  For the
present paper, however, my main interest in in the role of spin-orbit
coupling on the LTO phase, so the LTT phase will not be discussed in as great
detail.

\subsection{Splitting of the vHs}

\par
The nature of the splitting of the vHs depends sensitively on the magnitude of
the O-O hopping parameter, $t_{OO}$.  This is illustrated for four different
values of $t_{OO}$ in Figs. 4 (energy dispersion), 5 (density of states), 6
(Fermi surfaces) and 7 (electron filling).
The peaks in the dos correspond to vHs's, at which the topology of the Fermi
surface changes.  There can be up to {\it four} of these peaks, at different
energies, in the LTO phase.  In this section, the origin of each peak will be
discussed.  Note also that for $t_{OO}\le 0.1$, the dos vanishes at a point
between the vHs's.  This is a signal for a {\it diabolical point} in the energy
bands, which will be discussed in more detail below.
\par
It is easiest to begin with $t_{OO}$=0, Figs. 4a, 5a, 6a. In this case, there
is
an approximate electron-hole symmetry, so doping only up to half filling need
be
discussed.  For a nearly empty band, the first electrons go into pockets near
the $\Gamma$ and $M$ points (all descriptions are based on the tetragonal zone
-- in the orthorhombic zone, these points are equivalent).  With increased
doping, these pockets fill up and expand toward the diagonal, $\bar c_x+\bar
c_y
=0$, just as in the case without spin-orbit coupling.  Near half filling,
however, a first topology change occurs when small pockets appear near the $X$
and $Y$ points of the zone.  This is shown in detail in Fig. 8a.  At this
point,
all of the Fermi surfaces are closed and electron-like. Additional doping
causes all surfaces to grow further, until the two surfaces meet to produce
a rectangular box-like hole Fermi surface.  This electron-hole crossover marks
the point of the largest dos peak -- the main vHs.  With additional doping,
there is a single hole-like Fermi surface (in this quadrant of the Brillouin
zone), centered on the point $\bar M$.  As doping increases, the area of the
Fermi surface shrinks, until at half filling it is reduced to a single point
at $\bar M$.  This is the diabolical point: here, the spin-orbit
coupling is unable to lift the degeneracy of the two bands, leading to a
biconic structure, or `diabolo', in the energy surface.  At this point, the
dos vanishes.  For doping beyond half filling, the entire sequence is played
out
in reverse order. Fig. 7 illustrates the integrated electron density $n$
directly.  Since this is the integral of the dos, the steep rises correspond to
the vHs peaks, while the flat region is associated with the low dos regime near
the diabolical point.
\par
For finite $t_{OO}$, the doping dependence is similar, but more complicated,
since there is no longer electron-hole symmetry: the diabolic point is shifted
above the midway point between the two principal vHs's.  Thus, in Figs. 6b-d,
the topology changes are all similar for doping up to the lower main vHs,
corresponding to the box-like or coffin-like Fermi surface (dotted lines).
This
is presumably the Fermi surface of main interest for superconductivity, since
it
falls in the doping regime $x\sim 0.02-0.25$, depending on the value of $t_
{OO}$.  Note also that increasing $t_{OO}$ decreases the aspect ratio of the
box, leading to a somewhat squarer Fermi surface, with the corners moving away
from the $X$ and $Y$ points (see also Fig. 8b).  The long sides acquire some
curvature, giving the Fermi surface a coffin-like appearance.
\par
New features arise in the more electron-doped regime.  For $t_{OO}=0.1$, Figs.
4b, 5b, 6b, doping beyond the box vHs leads first to the diabolical point, and
hence a zero in the dos, just as for $t_{OO}=0$.  Beyond this point, however,
a new topology change occurs, as electron-like Fermi surface pockets open up
near the $X$ and $Y$ points (dashed lines).  Finally the main vHs on the
electron-doping side arises when these corner pockets intersect the growing
electron pocket centered on $\bar M$ (solid lines).
\par
Increasing $t_{OO}$ further, a point is reached when the electron pockets at
$X$
and $Y$ appear before the diabolical point is reached (Fig. 6c, $t_{OO}=0.25$).
Beyond this point, the Fermi surface becomes semi-metallic, with electron
pockets near $X$ and $Y$, and a hole pocket near $\bar M$.  The dos no longer
vanishes, and half filling corresponds to the doping at which the hole and
electron pockets have equal areas.  Doping beyond half filling, the hole pocket
shrinks to a point (the remnant of the diabolical point) (dot-dashed lines),
then an electron pocket grows from $\bar M$, until the electron pockets merge
at
the final vHs.  The case for $t_{OO}=0.65$ is similar.  Fig. 8b shows how the
box-like Fermi surface evolves with increasing $t_{OO}$.
\par
A similar series of topological changes arises in the LTT phase, Fig. 9.

\bigskip
\section{Diabolical Points of Energy Bands}

\bigskip\par
Whereas quantum mechanical energy levels tend to repel one another, to avoid
degeneracies, level crossings can appear for sufficiently complex Hamiltonians.
The analysis of such level crossings in molecules is an important aspect of
understanding reaction rates of chemical transitions.  For a diatomic atom the
degeneracy of two electronic levels is extremely rare, but Teller
showed that level crossings should regularly occur for larger, polyatomic
molecules\cite{Tell}.  Such crossings are closely related to the JT effect, and
Herzberg and Longuet-Higgins\cite{HLH} demonstrated that the degeneracy
leads to anomalous statistics: to make the electronic wave function single
valued, the {\it orbital} angular momentum must be quantized in half-integral
multiples of $\hbar$.  An alternative approach is to introduce a pseudo
magnetic field, so that single-valuedness of the electron is guaranteed by
associating a quantized vortex of pseudoflux at the molecular site\cite{pseud}.
\par
The degeneracy of the atomic levels leads to a linear dispersion of energy away
from the point of degeneracy, so that the energy surface has a biconical form
-- Berry has called such points {\it diabolical points}\cite{diab}.  A
signature
of such points is that the electronic phase changes by $\pi$ when the
electronic
orbit traverses a path in energy around the point\cite{LH}.  This phase change
was an early example of what is now known as Berry's phase\cite{Berry}.
Diabolical points have also been found in generalized two-dimensional Hubbard
models, in the presence of orbital antiferromagnetism or spin nematic
order\cite{Sch}, and similar phenomena
are found in a number of elemental semiconductors, notably
tin, which is a zero-gap semiconductor, and (approximately) graphite.  A key
parameter is the critical value of $t_{OO}$, at which the half filled band
undergoes a semimetal (SM) -- zero-band-gap-semiconductor (ZGS) transition.
This will be calculated as a function of $\lambda_i$ below.
\par
The Fermi surfaces near a diabolical point have many of the right properties to
explain a number of anomalous properties found in the cuprates, as summarized
in
the Discussion Section.  For example, near the diabolical point, the LTO phase
is stabilized by the umklapp splitting of the vHs; doping, if accomplished by a
rigid band shift, leads to hole-pocket Fermi surfaces, which provide the
correct
Hall density; and finally, sufficient doping
repositions the Fermi level at the new vHs of the box/coffin-like Fermi
surface, close to the doping of optimum T$_c$.
\par
The form of two-fold degeneracy illustrated in Figs. 1a, 2a, is not unique to
the LTO phase of LSCO, but is a common feature of a number of space groups
containing glide planes.  In all of these cases, spin-orbit coupling lifts
most of this degeneracy, leaving a residual two-fold degeneracy on a few
isolated points or lines.  In the case of LSCO, when the LTO energy bands are
treated as two-dimensional, the residual degeneracy is a clear example of a
diabolical point, as illustrated in Figs. 4a,b.  The situation can be somewhat
more complex in three dimensions.  Thus, for LSCO, when c-axis dispersion is
included, there is an isolated degeneracy (diabolical point) at one point along
the line perpendicular to $Y^*$, while the entire line passing through $X^*$
retains a two-fold degeneracy.  Since this dispersion is small, its inclusion
should not greatly change the results of the present paper.
\par
The diabolical point is a {\it novel form of vHs}, associated with overlapping
bands.  This can most easily be understood with reference to the dos, Fig. 5.
In Fig. 5a,b the dos goes to zero at the diabolical point.  This is the
characteristic form of the dos at the top or bottom of a band, so the
diabolical
point can be understood as the point where one band terminates and another
immediately begins.  Note that the electron density $n$ is defined as the
number
of electrons per Cu; since there are two Cu per unit cell in the LTO phase, the
diabolical point appears when there are exactly two electrons per cell -- an
exactly filled band.  However, when $t_{OO}$ increases, the topological change
occurs while the two bands still overlap in other parts of the Brillouin zone,
Figs. 5c,d.  In this case, the diabolical point is shifted away from half
filling.  The local contribution to the dos from the region of
the diabolical point vanishes, but there is a finite total dos due to the
contribution from other parts of the Fermi surface.
\par

\bigskip
\section{Slave Boson Calculations: Choice of Parameters}

\subsection{Spin Corrected Slave Boson Theory}

\par
Slave boson calculations offer a systematic procedure for incorporating
correlations into a tight-binding model of the CuO$_2$ planes.  The model has a
number of desirable features, including renormalization of the bandwidth near
half filling, and a metal-insulator (Mott) transition at exactly half filling
above a critical value of the ratio $\Delta /t_{CuO}$.  A key limitation is
that
the theory is formally a large-N theory, where N is the degeneracy of the
electron states on a Cu site, and the calculation is usually carried out only
to
lowest order (in $1/N$).  At this order, the slave boson calculation does not
include magnetic effects.  Moreover, $N\simeq 2$ for the cuprates, so it is
doubtful whether the lowest-order predictions of the theory are quantitatively
correct.  Nevertheless, detailed calculations are important -- for example,
to see how close the cuprates are to the metal-insulator transition, for a
`realistic' choice of parameters.  In the present context, it is also important
to determine how close the state at half filling is to the diabolical point.
\par
In the large-N calculations, some parameters are renormalized by a factor $N$,
which then does not explicitly appear in the final Hamiltonian.  Nevertheless,
in choosing the `correct' parameters to use in the calculations, this factor
should be included.  In order to minimize confusion, I will write the `large-N'
parameter as primed, and the bare parameter as unprimed.  In particular, in
comparing the present results to e.g., Ref. \cite {Rai}, one should take $t_{
CuO}^{\prime}=\sqrt{2}t_{CuO}$, $V^{\prime}=2V$, where $V$ is the
nearest-neighbor Coulomb repulsion.
\par
In conventional slave boson calculations, the factor $N$ includes both spin
and orbital degeneracy (for the cuprates, $N=2$ due to spin
degeneracy\cite{KLR}).  However, I believe that it is {\it not} appropriate to
include the spin degeneracy in the term $t_{CuO}^{\prime}$.  The factor $N$
arises as follows: if the electron can hop from the Cu to $N$ different
orbitals
on a given O, and all of these orbitals have the same energy and hopping $t$,
then the antibonding band of the $N+1\times N+1$ matrix has the same form as
that of a $2\times 2$ matrix, with a single effective O orbital, if the hopping
energy is renormalized to $\sqrt{N}t$.  Such an enhancement would arise in
the cuprates only if spin flip hopping were as probable as spin-non-flip
hopping.  But the spin-flip hopping is considerably less probable, and in fact
is explicitly calculated here as spin-orbit coupling.
\par
If spin-flip and non-spin-flip scattering are equally probable, then the
electron loses spin memory on each hop, greatly modifying the magnetic
properties of the model.  Hence, I propose a `spin-corrected' version of slave
boson theory, in which spin-flip and non-spin-flip scattering are explicitly
distinguished.  In this case, the bare hopping parameter $t_{CuO}^{\prime}$
becomes
$$t_{CuO}^{\prime}=\sqrt{{N\over 2}}t_{CuO}.$$
Since all bare parameters are normalized to $t_{CuO}$, this modification does
not directly affect any of the predictions of the model.  However, it is
crucial
when one attempts to determine whether the Mott transition occurs in the
cuprates, by estimating `realistic' values for the parameters.  For instance,
the Mott transition occurs only if $\Delta$ is larger than a critical value,
$\sim 3.35 t_{CuO}^{\prime}$.  For realistic values of $\Delta$, a transition
will occur only if the factor $\sqrt{N}$ is not included.
\par
For convenience in the subsequent discussion, I will carefully distinguish the
two cases, labelling the spin-corrected parameters with a superscript $t$
(i.e., when it is assumed that $t_{CuO}^{\prime}=t_{CuO}$), and without the
superscript for the conventional slave boson theory (assuming $t_{CuO}^{\prime}
=\sqrt{2}t_{CuO}$).
Thus, when $\Delta_0=6\sqrt{2}eV=8.5eV$, then $\Delta_0^t=6eV$.
\par
In contrast, the nearest-neighbor Coulomb repulsion $V$ is the same for both
spins, so the factor $N=2$ should remain for $V^{\prime}$.
\par
In searching through the literature, I was unable to find any explicit
discussion of the proper choice of degeneracy factors.  However, it is clear
that different groups have made different choices, and that this choice is a
cause for significant variations in the resulting slave boson calculations.  To
be specific, let $b$ be the slave boson amplitude, and $r_0$ the
renormalization
of the hopping parameter.  Then $b$ is defined through the equation
$$\sum_{i=1}^Nd^{\dagger}_id_i+b^{\dagger}b=q_0N,$$
where $N$ is the electron degeneracy and $q_0$ is the filling factor (for the
present case, $N=2$, $q_0=1/2$), while $r_0$ is defined by $t_R=r_0t_{CuO}$.
Given these definitions, then previous slave boson calculations have used a
variety of values for the ratio $\eta_b =r_0/b$, with Refs. \cite{SH,New,QSi}
assuming $\eta_b=1$, Refs. \cite{KLR,Cast} assuming $\eta_b=1/\sqrt{2}$, and
Refs. \cite{Gri,Rai} (apparently) assuming $\eta_b=1/2$ (since Ref. \cite{Rai}
uses the same symbol for $t_{CuO}$ and $t_{CuO}^{\prime}$, it is sometimes
difficult to ascertain which is meant).  In general, most of the calculations
assuming $\eta_b=1$ find that the metal-insulator transition does not occur at
half filling, for reasonable parameter values, while the transition is found if
a smaller $\eta_b$ is assumed.  The present calculations confirm {\it both}
these results.

\subsection{Bare Parameter Values: `Large' Parameters}

\par
While the technique of carrying out the slave boson calculations to lowest
order
in $1/N$ is reasonably well understood, and considerable progress has been made
in calculating the next-order corrections, relatively little attention has been
paid to the proper choice of the starting, bare parameters.  This choice is
crucial, particularly in the immediate vicinity of half filling, where it
controls whether or not a metal-insulator transition will occur.  Under the
present circumstances, the introduction of additional parameters -- to describe
spin-orbit and electron-phonon coupling -- renders a detailed discussion of the
choice of parameters even more urgent.  Two separate problems arise: how to
choose the bare parameters, and how the parameters vary with doping.  This
section will attempt to provide an introduction to the problems involved, but
cannot pretend to provide a definitive solution.
\par
The usual technique which has been adopted is to extract the bare parameter
values from an LDA or cluster calculation of the band structure.  However,
simply parametrizing the the LDA-derived Fermi surfaces is {\it not}
appropriate, since these calculations already include
correlation effects in an average way.  Fortunately, for LSCO a number of
groups have analyzed the band structure calculations, to provide estimates for
the bare parameters.  Table I lists the values produced by several groups,
using
LDA or cluster calculations\cite{Hyb,GraM,Esk}.  Earlier calculations are
summarized in Ref. \cite{MAM}.
\par
Even knowing these values is not sufficient: it must be regognized that the
three-band model is a simplification of the physics, and the parameters which
enter it are {\it effective} parameters, which describe the combined effect of
several microscopic parameters.  Two examples are of particular importance.
\par
The nearest neighbor Coulomb repulsion $V$ acts to renormalize the bare Cu-O
separation\cite{BARA}, $\Delta_0$, according to\cite{Rai}
$$\Delta_0\rightarrow\Delta_0+2V(1-x-4r_0^2) ,\eqno(7)$$
where $r_0$ is the mean field amplitude of the slave boson.  While the term
$2V(1-x)$ is just a constant, and can be absorbed into an x-dependent
$\Delta_0$,
the last term depends self-consistently on the renormalization, and hence gives
rise to a possible instability against phase separation\cite{Rai}.  However,
the
critical value of $V$ required for this phase separation ($\sim 1.76t_{CuO}^{
\prime}/2\sim 1.6eV$\cite{Rai} in the standard slave boson theory, or $\sim 1.6
/\sqrt{2}=1.1eV$ in the spin-corrected version) appears to be somewhat larger
than the value estimated from LDA-type analyses, and moreover near half filling
the value of $r_0^2$ tends to be very small.  Hence, in the present
calculations
Eq. 7 will be approximated by enhancing the effective value of $\Delta_0$.
\par
The second example involves the proper choice of the effective O-O hopping
parameter.  The LDA-derived value of $t_{OO}$ is too large to
give a good description of the curvature of the Fermi surfaces, suggesting that
higher-lying bands play an important role in modifying the Fermi surface
shape\cite{LZAM}. Aligia\cite{AAA} has shown that this parameter is the sum of
two effects of opposite sign: the direct O-O hopping, of value estimated in
Table I, and the Cu $d_{x^2-y^2}-d_{z^2}$ interference term, which acts to
reduce the curvature.
\par
The problem of calculating the effective value for $t_{OO}$ is similar to
the problem of calculating $t^{\prime}$ in a $t-t^{\prime}-J$
model, since the role of both parameters is to distort the Fermi surface away
from square at half filling. In this case, Jefferson, et al.\cite{JEF}, have
shown that even within a three-band model correlation effects can act to reduce
$t^{\prime}$ toward zero, and can even change its sign.
\par
In Appendix II, I derive an effective value of $t_{OO}$ from a five-band model
of the CuO$_2$ planes, including the Cu $d_{z^2}$ and apical O
orbitals\cite{FGD}.  A substantial reduction is found, and $t_{OO}$ could even
change sign.  Due to uncertainties in many of the parameter values, it is
difficult to use this calculation to fix the correct effective value for
$t_{OO}$.  Hence, I propose an approximate criterion for an acceptible value of
$t_{OO}$: that the vHs fall at the same doping as in the LDA calculations, for
the self-consistently renormalized slave boson values.
The rationale for this is that correlation
effects will be smallest away from half filling, so that the approximate manner
in which LDA calculations incorporate correlation effects are likely to be most
nearly correct near the vHs.  The remarkable degree to which these calculations
reproduce the experimentally measured Fermi surfaces confirm the plausibility
of
this choice.  This estimate of the effective value of $t_{OO}$ is in general
agreement with the reduction estimated in Appendix II.
\par
Figure 10 shows how $x_{vHs}$, the doping at which the Fermi level coincides
with the vHs, varies with $t_{OO}$, assuming bare values of $t_{CuO}=1.3eV$ and
$\Delta$ = 4 or 6eV.  Using the bare parameters, the vHs falls in an acceptable
regime (I estimate\cite{vHs} $x_{vHs}$ = 0.16 in LSCO and 0.25 in YBCO).
However, correlation effects greatly reduce the value of $t_{CuO}$, and hence
enhance the ratio $t_{OO}/t_{CuO}$, which controls the Fermi surface curvature.
Thus, to recover reasonable values for $x_{vHs}$, it is necessary to assume
smaller values for $t_{OO}$.  For definiteness, I have assumed bare values of
$\Delta_0^t=4eV$ or $6eV$, in which case $t_{OO}^t\simeq
0.14-0.2eV$ for LSCO, and 0.3-0.4eV for YBCO.
\par
It is tempting to speculate that the lower $T_c$ in LSCO is related to the
finding that the effective $t_{OO}$ is only half as large in LSCO as in YBCO.
A smaller $t_{OO}$ can reduce $T_c$ directly, because the excess hole
population
is smaller at the vHs, or indirectly, since a smaller $t_{OO}$ means a more
nearly square Fermi surface at half filling, thereby enhancing the possibility
of a structural instability which competes with superconductivity.  If $t_{OO}$
is lowered by repulsion from the $d_{z^2}$ level, then this parameter is
particularly sensitive to the environment off of the CuO$_2$ planes -- e.g.,
other layers, and the apical oxygen.  Indeed, there is a close sorrelation
between $\Delta_d$, the $d_{x^2-y^2}-d_{z^2}$ splitting, and the distance $d_A$
between the Cu and the apical oxygen (Appendix III).  This correlation can be
interpreted in terms of a static JT-like effect (N.B.: not to be confused with
the vHs-JT effect) which produces the distortion of the six O's surrounding the
Cu from a perfect octahedron\cite{JT,Pick}.  Hence, anything which reduces
the Cu-apical O separation also reduces the $d-d$ splitting, in turn reducing
the effective value of $t_{OO}$.  The correlation between small T$_c$ and small
optimum hole doping has been known for a long time, and it is also known that
the apical oxygen plays some role, since a larger apical O potential correlates
with higher T$_c$\cite{Ohta}.

\subsection{Bare Parameter Values: Spin-Orbit and Phonon Coupling}

An important role in the theory is played by the magnitude of the spin-orbit
coupling.  Specifically, if $\lambda\equiv\lambda_1+\lambda_2$ is large enough,
the state at half filling will be a zero-gap semiconductor (ZGS) rather than a
semimetal.  The location of the
zero-gap semiconductor -- semimetal (ZS) phase boundary depends sensitively on
the relative magnitudes of $\lambda$ and $t_{OO}$.  For $\lambda =50meV$, the
ZS crossover occurs at $t_{OO}\simeq 0.13eV$; for $\lambda =6meV$, the
crossover
occurs for $t_{OO}$ about an order of magnitude smaller.
\par
The parameter $\lambda$ has been estimated by Bonesteel, et al.\cite{Bones}, as
$$\lambda_2\simeq ({\Delta g\over g})t_{CuO}\theta,\eqno(8)$$
where $\theta$ is the octahedral tilt angle, and $\Delta g$ is the shift in the
electronic $g$-factor $g$ due to spin-orbit coupling.  With the estimates
$\Delta g/g\sim 0.1$\cite{Thio}, $t_{CuO}=1.3eV$ (I use the bare value of
$t_{CuO}$ to estimate the bare value of $\lambda$), and experimental values for
$\theta$, Eq. 8 gives $\lambda_{LTO}\simeq 6meV$, $\lambda_{LTT}\sim 24meV$.
\par
There are a number of factors which could substantially enhance the value of
$\lambda_{LTO}$.  First, if there is a dynamic JT effect, the experimentally
observed value of $\theta$ will be smaller than the microscopic local values,
which would be more representative of local LTT-like distortions.  Secondly,
dynamical JT effects can greatly reduce the observed spin-orbit corrections to
$g$-values (the Ham effect\cite{Ham}).  Thirdly, $\lambda$ is playing a role
similar to umklapp scattering in splitting the vHs degeneracy, and in the
analogous one-dimensional problem, it is well known that renormalization
effects
greatly enhance an initially small umklapp term (for a discussion of the
connection between the vHs problem and one-dimensional umklapp scattering, see
\cite{Recon}).  Finally, as will be seen in the following subsection, what
is important in the present problem is how the parameters renormalize in the
presence of strong correlation effects.  I am assuming that $\lambda$ is
renormalized in such a way that $\lambda /t_{CuO}$ remains constant.  This
assumption is by no means self-evident -- Bonesteel, et al.\cite{Bones},
suggest that $\lambda$ is unaffected by strong coupling.  From the perspective
of the present analysis, any increase in the ratio $\lambda /t_{CuO}$ is
equivalent to starting with a larger effective bare value of $\lambda$.  For
all these reasons, it is plausible to assume that $\lambda_{LTO}$ is large
enough that the state at half filling
is close to the ZGS phase.  Clearly, a detailed microscopic estimation of the
parameters, particularly $\lambda$ and $t_{OO}$ is a desideratum.
\par
The electron-phonon coupling parameter can be estimated as\cite{RM8A,RM8}
$$\eta ={2\beta_{\pi}^2t_{CuO}^2\theta^2\over\Delta}\simeq 2.16\theta^2 eV,
\eqno(9)$$
where $\beta_{\pi}\simeq 1.6$.  This is close to the recent estimate of Song
and
Annett\cite{SongAn}, $\eta\simeq 2.8\theta^2 eV$.  Using $\theta_{LTO}\simeq
0.5$\cite{Bones}, $\eta_{LTO}\simeq 5.4meV$.  In the LTT phase, $\theta_{LTT}$
is about four times larger, yielding $\eta_{LTT}\simeq 86meV$.  The latter
value
would also be approximately correct for a dynamic JT effect in the LTO phase.

\subsection{Doping Dependence: Scaling with $t_{CuO}$}

\par
In the absence of spin-orbit coupling and electron-phonon interaction, the
slave
boson theory provides a system of self-consistent equations for determining how
$\Delta$ and $t_{CuO}$ vary with doping, due to correlation effects (Eqs. 11,
below).  An important question is how $t_{OO}$, and, in an extended theory,
$\lambda$ and $\eta$ vary with doping.  This question can be conveniently
separated into two parts: since strong correlation effects can renormalize
$t_{CuO}$ to zero, while $\Delta$ remains finite, one must first ascertain how
the parameters scale with $t_{CuO}$, and then check if there are residual
finite renormalizations, as with $\Delta$.
\par
The scaling can be determined from the theoretical expressions for the
parameters,
$$\lambda\sim {\lambda_0 t_z\over\Delta_d},\eqno(10)$$
and Eq. 6b for $\eta$.  Here $\lambda_0$ is the spin-orbit coupling
constant\cite{Coff}, $t_z$ is the hopping parameter between the planar O
$p\sigma$ orbitals and the Cu $d_{z^2}$ orbital, and $\Delta_d$ is the
Cu $d_{z^2}-d_{x^2-y^2}$ splitting energy.  Now correlation effects inhibit the
placing of two holes on the same Cu, and this causes the renormalization of
$t_{CuO}$ to zero at half filling.  Thus, $t^{\prime}$ should also scale to
zero, since it involves the same Cu orbital.  The case with $t_z$ is slightly
different, since it involves a different Cu orbital, but the Hubbard $U$ for
this orbital is also large, and the simplest assumption is that $t_z$ also
scales to zero at half filling.
\par
Hence, the following assumption will be made for the scaling of the parameters.
First, $\lambda$ is renormalized proportionally to $t_{CuO}$, while $\eta$
scales as $t_{CuO}^2$ (!)  Finally, $t_{OO}$ should not rescale with $t_{CuO}$.
\par
Because of the strong rescaling of $\eta$, it might be thought that this
electron-phonon coupling could be neglected near half filling, in comparison
with the electron-phonon coupling term, $\lambda$.  However, as the scaling
analysis of Section VB shows, the gaps produced by both terms scale in the
same fashion near half-filling.  Because of the different ways they enter the
Hamiltonian, Eqs. 2 and 6, the effective electron-phonon interaction strengths
are $g\sim\eta$ and $g^{so}\sim\lambda t_{CuO}/\Delta$, both of which scale as
$g\sim r_0^2$ near half filling.  The electron-phonon coupling parameter has a
similar scaling, $\lambda_{ep}\sim g^2N(E_F)\sim r_0^2$, since $N(E_F)$, the
dos
at the Fermi level, varies as $N(E_F)\sim 1/t_{CuO}^2$.  A similar scaling has
been found in a number of earlier studies\cite{KiT}.  I believe that this
suppression of electron-phonon coupling by correlations is an artifact of the
mean-field theory.  Anderson\cite{And} has found that correlations actually
enhance nesting effects, and Eliashberg\cite{Eli} finds that electron-phonon
coupling can remain strong in the presence of strong correlation effects.
This issue will be discussed further in a future article.
\par
A separate issue is the doping dependence of the LTO transition temperature,
which drops to zero around $x\simeq 0.2$.  Since $\lambda\sim\theta$ and $\eta
\sim\theta^2$, these parameters will suffer a further reduction with doping.
Ultimately, of course, the LTO transition temperature should be calculated
self consistently in the presence of correlation effects.

\subsection{Residual Doping Dependence}

\par
In addition to the strong rescaling noted above, there can also be finite
renormalizations with doping, similar to the renormalization of $\Delta$.  As
will be shown below, the renormalized $\Delta$ is typically a factor of three
smaller than its bare value.  This renormalization of $\Delta$ will have an
impact on several of the other parameters.  For instance, $\eta\propto\Delta^
{-1}$ will be increased by the same factor of three.  The effect on other
parameters depends on how $\Delta_d$ varies with doping.  Since $\Delta_d<<
\Delta$, it is necessary for the $d_{z^2}$ level to shift in tandem with the
$d_{x^2-y^2}$ level, to keep the states at the Fermi level of $d_{x^2-y^2}$
symmetry.  However, in this shift, it is not clear whether the two levels
maintain a constant separation, or whether $\Delta_d$ is itself reduced
proportional to $\Delta$.  The latter situation arises naturally when a single
slave boson is introduced for all d-levels, and has been assumed in a recent
study of the role of the apical oxygens\cite{FGD}  In this case,
$\lambda\propto
\Delta_d^{-1}$ would be further increased by the same factor.
\par
A similar effect would arise for the effective value of $t_{OO}$.  While the
microscopic $t_{OO}$ is unlikely to be strongly affected by Cu Coulomb
correlations (but see \cite{JEF}), the counterterm, due to repulsion from the
$d_{z^2}$ levels, will be enhanced by reducing $\Delta_d$.  The form of the
reduction is suggested in Table II (where an exlpicit form for $X$ is given in
Appendix II).
\par
In summary, there is much left to be understood about the proper choice of
parameter values, and how they change in the presence of strong correlations.
The fact that reducing $\Delta$ increases some other parameters could even lead
to an instability, just as occurs with sufficiently large values of $V$, Eq. 7.
Lacking more precise knowledge, the following prescription will be
followed here.  Only renormalization of $\Delta$ and $t_{CuO}$ will be
explicitly accounted for in the slave boson calculations.  It will, however,
be assumed that $\lambda$ scales with $t_{CuO}$, and $\eta$ with $t_{CuO}^2$.
This will be shown to lead to an interesting scaling regime when $t_{CuO}$ is
sufficiently small.  The other, minor renormalizations discussed in the present
subsection will be approximately accounted for by adjusting the values of the
bare parameters.  Since the renormalization is strongest near half filling,
the bare parameters will be taken as those appropriate to half filling, and
effects due to their variation with doping will be neglected.  A summary of the
present parameter estimates is given in Table II.
\bigskip
\section{Slave Boson Calculations in the Three Band Model}

\subsection{Slave Boson Technique}

The strong on-site Coulomb repulsion acts to greatly reduce the probability of
double occupancy of an atomic orbital.  The effect is most important for Cu,
since $U$ is larger for copper and since the Cu band is very close to
half filled, so the probability of finding an unoccupied Cu site is small.
Hence, for simplicity $U$ will be neglected on the O sites, and set to infinity
on Cu -- this has the effect of forbidding any double occupancy of the Cu's.
In a slave boson calculation, this constraint is enforced by introducing a
boson to occupy the second orbital of each singly occupied site.  By satisfying
the constraint in mean field, it is possible to reduce the calculation to a
self-consistent renormalization of the bands.  Basically, the Cu-O hopping
parameter is reduced by the renormalization, while $\Delta$ tends to increase,
as the antibonding band becomes more purely Cu-like.
\par
The equations of self-consistency can be written
$$r_0^2={1\over 2}[1-\sum_ku_k^2f_h(E_k)],\eqno(11a)$$
$$\Delta_0-\Delta ={1\over 2r_0^2}\sum_ku_k^2f_h(E_k)(E_k-\Delta ),\eqno(11b)$$
where the renormalized parameters are $t_R=r_0t_{CuO}^{\prime}$ and $\Delta$,
$f_h(E_k)$ is the Fermi function, and $u_k$ is the d-wave amplitude of the
wave function.  $E_k$ and $u_k$ are calculated using the renormalized
parameters, and it is assumed that $t_{OO}$ is not renormalized.  Only the case
$T=0$ will be considered in the present paper.  As written, Eqs. 11 are valid
for a hole picture, so $f_h(E_k)=1$ for $E_k>E_F$, and $=0$ otherwise.  There
is an equivalent electron picture, discussed in Appendix IV.  The sums are
normalized per Cu atom, so for example
$$\sum_k(1)=6,\eqno(12a)$$
that is, that there are three bands (spin 1/2), and
$$\sum_kf_h(E_k)=1+x.\eqno(12b)$$
\par
Equations 11 are written in such a way as to remain valid in the presence of
spin-orbit coupling and electron-phonon coupling.  These equations are solved
self-consistently by guessing initial values for the
renormalized values and numerically performing the double integrals to allow
calculation of the corresponding bare values, then readjusting the initial
guesses until the known values of the bare parameters are recovered.  [This
improves upon my earlier work\cite{RM3}, which approximated the problem by a
single integral.]  If $t_{R}$ is small enough, however, the equations simplify
to a scaling form, for which only a single parameter, $\Delta$, need be varied.
This scaling regime is discussed in the following Subsection.

\subsection{Scaling Regime}

\par
When $t_{R}\sim 0$, and $\Delta >>4t_{OO}$, the antibonding band is very
narrow, $\delta\equiv E-\Delta <<E$.  In this case, the dispersion relations,
such as Eq. 4, can be solved to lowest order in $\delta$, setting $E
\rightarrow \Delta$ in all the remaining terms.  The resulting solution
posesses a simple scaling form.
\par
It is convenient to first treat the special case, $\lambda =\eta =0$ -- the
untilted case.  In this case, the normalized dispersion $\gamma =\delta /\hat
t$
(with $\hat t=4t_R^2/\Delta$) depends only on the parameter $y=4t_{OO}/
\Delta$:
$$\gamma ={s_x^2+s_y^2+2ys_x^2s_y^2\over 1-y^2s_x^2s_y^2}.\eqno(13)$$
The resulting bandwidth is $2\hat t/(1-y)$, and the band is nearly pure
Cu-like,
with
$$u_k^2=1-{\epsilon\hat t\over\Delta},\eqno(14a)$$
$$\epsilon ={1\over (1-ys_xs_y)^2}\bigl[s_x^2+s_y^2-{2ys_xs_y(s_x-s_y)^2\over
(1+ys_xs_y)^2}\bigr].\eqno(14b)$$
Letting $\bar\gamma =\sum_k\gamma f_h(E_k)$ be the average of $\gamma$ over the
occupied states, with a similar definition of $\bar\epsilon$, the
self-consistancy equations become
$$\Delta_0=\Delta +{4t_{CuO}^{\prime 2}\bar\gamma\over\Delta},\eqno(15a)$$
$$t_R^2={x\over {4\bar\epsilon\over\Delta^2}-{2\over t_{CuO}^{\prime 2}}},
\eqno(15b)$$
where $x$ is the hole doping.  Note that once $\Delta$ is known, Eq. 15b is an
explicit equation for $t_R$, and that in the scaling regime $t_R\rightarrow 0$
at half filling.
\par
Remarkably, using the scaling of the small parameters discussed above, the
same scaling regime holds for the full dispersion relation, Eq. 4.  Thus, Eq.
A1 in Appendix I is a quadratic equation for $\delta$, and it can easily be
seen that when $\lambda$ scales $\propto t_{R}$, and $\eta\propto t_{R}^2$,
then $\delta\propto\hat t$, as found above.  Now, however, the energy
dispersion
is a function of three variables $y$, $\lambda /t_R$, and $\eta /t_R^2$.
The bare parameter values are sufficiently close to this scaling regime that
most of the results of Figs. 1-9 are only qualitatively changed if the
renormalized parameters are used in place of the bare ones (except when $t_{OO}
=0.65eV$, for which $4t_{OO}<<\Delta$ is not always satisfied).

\subsection{Slave Boson Results: (a) Strong Correlation
($t\rightarrow 0$) Limit}

\par
The above scaling results can be used to analyze the strong correlation limit
-- specifically, to explore the circumstances in which $t_R$ is
renormalized to zero.  In particular, Eq. 15b shows that {\it $t_R=0$ can
only occur at half filling, $x=0$}.  This also follows directly from Eq. 11a
when $r_0=0$ and $u_k=1$, and is consistent with the result for the
one-dimensional Hubbard model\cite{LW}.
\par
Equation 15a shows that there is a {\it minimum value of $\Delta_0$} for which
$t_R=0$ is possible.  Defining $\Delta_m^2=4t_{CuO}^{\prime 2}\bar\gamma_0$,
where $\bar\gamma_0$ is $\bar\gamma$ evaluated for a half filled band, then the
minimum value of $\Delta$ is $\Delta_m$, and at this point
$\Delta_0=2\Delta_m$.
Numerically evaluating $\bar\gamma_0$ as a function of $y$ thus provides a plot
of $\Delta_0/t_{CuO}^{\prime}$ vs $t_{OO}/t_{CuO}=y\Delta_m/4t_{CuO}^{\prime}$,
Fig. 11.
For $y=0$, $\bar\gamma_0$ can be evaluated analytically, yielding\cite{Cast}
$${\Delta_0\over t_{CuO}^{\prime}}=4\sqrt{{1\over 2}+{2\over\pi^2}}=3.353,
\eqno(16a)$$
or $\Delta_0^t=4.36eV$ when $t_{CuO}=1.3eV$.
Note that for the conventional theory, $\Delta_0=\sqrt{2}\times 4.36=6.17eV$.
If a larger value, $t_{CuO}=1.4eV$, is assumed, then $\Delta_0=6.64eV$,
in good agreement with the 6.5eV found by {Sudb\o} and Houghton\cite{SH}.
\par
The above result is, however, only a lower limit to the critical $\Delta_0$
when
$t_{OO}\ne 0$.  The analysis assumes that the system is in the scaling regime,
which is not necessarily the case when $t_{OO}$ is comparable to $\Delta /4$.
A more accurate phase boundary can be found from numerically solving the
self-consistent Eqs. 11, and this is also illustrated in Fig. 11.  Care must be
taken in deriving these points, since the numerical calculations become
unstable
when $t_R\sim 0$.  The procedure followed here is to choose a value for the
renormalized $t_R$ and adjust $\Delta$ until $t_{CuO}=1.3eV$.  A number of
values of $t_R$ were chosen in the range 0.075-0.2eV, and it was found that
a plot of $\Delta_0$ vs $t_R^2$ gave a straight line, which was extrapolated to
$t_R=0$ to generate the points plotted in Fig. 11 as open circles.  These
closely approximate a straight line
$$\Delta_0=3.353t_{CuO}^{\prime}+2.94t_{OO}.\eqno(16b)$$
\par
This metal-insulator phase diagram is in semi-quantitative agreement with
Monte Carlo calculations on the three band Hubbard model, at least at $t_{OO}=
0$, where most of the calculations have been done.  Assuming a large but finite
value for the on-site Coulomb repulsion, $U_d=6t_{CuO}$, a charge-transfer
insulating gap appears to open up for $\Delta /t_{CuO}$ between 1 and
4\cite{Dopf}, or between 2 and 3\cite{Scal}.  The resulting gap values are also
comparable.  This good agreement holds {\it for the spin-corrected theory},
while the conventional slave boson theory is off by a factor $\sim\sqrt{2}$.
\par
The ZS crossover can also be determined analytically.  The diabolical point
is the solution of Eq. A1 (Appendix I) at the $\bar M$ point ($\bar c_x=\bar
c_y
=0$), or
$$E_d\delta_d={2(\lambda_A^2+2t^2)+y^{\prime}(\lambda_B^2+8t^2)\over 1-4y
^{\prime 2}},\eqno(17a)$$
where the subscript $d$ refers to the diabolical point, the various $\lambda_
i$'s are defined in Appendix I, and $y^{\prime}=t_{OO}/E_d$.  Equation 17a is
true in general; in the scaling regime, each explicit $E_d$ should be replaced
by $\Delta$.  The ZGS state is present when there are no other solutions of
Eq. A1 along the diagonal $\bar c_x+\bar c_y=0$.  When such solutions exist,
the diabolical point falls in a semimetallic regime.  The crossover ooccurs
when
the second solution falls at the $X$ or $Y$ point, in which case
$$E_d\delta_d=2(\lambda_A^2+2t^2)\pm 4\sqrt{2}t\lambda.\eqno(17b)$$
The upper solution is relevant here; the lower one falls close to the box-like
Fermi surfaces.  Equating Eqs. 17a,b yields the equation of the crossover
$$y^2[\hat\lambda_1^2+\hat\lambda_2^2+\sqrt{2}(\hat\lambda_1+\hat\lambda_2)+1]
+y[2-\hat\lambda_1\hat\lambda_2]-4\sqrt{2}(\hat\lambda_1+\hat\lambda_2)=0,
\eqno(17c)$$
where $\hat\lambda_i=\lambda_i/t_{CuO}$.  The resulting crossover phase diagram
is shown in Fig. 12.

\subsection{Slave Boson Results: (b) Metal-Insulator Transition}

\par
The full Equations 11 have been solved for a variety of bare parameters $\Delta
_0$, $t_{CuO}$ and $t_{OO}$, as a function of hole doping $x$.  The following
results are found.  (1) For $x\ne 0$, the system is metallic, with $t_R>0$. (2)
There is a critical value $\Delta_c$ (Eq. 16), such that $t_R\rightarrow
0$ as $x\rightarrow 0$, if $\Delta_0 >\Delta_c$. (3) For $\Delta >\Delta_c$,
the
Fermi level jumps discontinuously as $x$ passes through zero, with the
discontinuity being the charge-transfer gap.
(4) The renormalization is such as to keep the Fermi level pinned near the vHs
-- the more so as $\Delta$ increases.  This pinning was found in earlier
slave boson\cite{RM3,New,QSi} and other\cite{Pines} calculations.
\par
Using the parameters estimated above (Table II), the present calculations
predict a metal-insulator transition at half filling, but only if $t{CuO}^
{\prime}=t_{CuO}$ (the parameters with superscript $t$ in Fig. 11). If, on the
other hand, it is assumed that $t{CuO}^{\prime}=\sqrt{2}t_{CuO}$, then the
half filled band remains metallic, $\Delta <\Delta_c$.

The metal-insulator transition is illustrated in Fig. 13, both in the presence
of spin-orbit coupling, $\lambda_2=0.05eV$ (symbols) and in its absence
(lines),
assuming $t_{OO}^t=0.25eV$, as $\Delta_0^t$ increases from $4eV$
(solid line or open squares) to $5eV$ (dashed line of filled circles)
to $6eV$ (dot-dashed line or open circles).  The transition in the
absence of spin-orbit coupling will be discussed first.  Figure 13a
shows that $t_R$ renormalizes to zero exactly at half filling ($\Delta_0^t=5
eV$ is just below the transition, so $t_R$ has a small but finite value
at half filling).  The scaling theory, Eq. 15b, approximates the doping
dependence of $t_R$ near half filling (dotted line). Near this point, the Fermi
energy changes rapidly (Fig. 13b), from a value near $\Delta_0$ (Cu-like
carriers) when $x<0$ to a much smaller value (more O-like) for $x>0$.  This
change is discontinuous above the Mott transition; this is the charge-transfer
insulator gap\cite{KLR}, and should be compared to the one-dimensional Hubbard
model (see Fig. 3 of Ref. \cite{Recon}).  For even larger $x$, $\Delta$ becomes
negative, and the Fermi level falls directly in the O-like band.  [In this
regime, the three-band model should not be trusted, because of the proximity of
the $d_{z^2}$ band to the Fermi level.  Indeed, in this doping regime an
enhanced $d_{z^2}$ character is often observed experimentally.]
\par
The quasi-pinning of the Fermi level to the vHs is illustrated in Figs. 13c,d.
In a rigid-band filling picture, the bandwidth and the position of the vHs
would both be independent of doping, so the separation between the vHs and
Fermi
level would smoothly track the variation of the Fermi level with doping.  This
is not what happens in the presence of strong correlation effects.  Figure 13c
shows that the position of the vHs (defined by the hole doping $x_{vHs}$ at
which the vHs would coincide with the Fermi level) actually changes with
doping,
while Fig. 13d shows the energy separation between the Fermi level and the
vHs, $\Delta E=E_F-E_{vHs}$.  Note for example the case $\Delta_0^t=6eV$:
over the doping range $x=-0.1$ to $x=+0.3$, the Fermi energy varies by $4
eV$, while $|\Delta E|\le 10meV$!  Thus, correlations strongly pin the
vHs near the Fermi level, especially when a Mott transition occurs at half
filling.
\par
Figure 14 illustrates the corresponding Fermi surfaces, both at the vHs (i.e.,
when $\Delta E=0$) and at half filling.  Note that these latter Fermi surfaces
are far from square, and their proximity to the vHs arises mainly from the very
small bandwidth.
\par
The changes in $t_R$ and $E_F$ due to inclusion of spin-orbit coupling are
relatively small, Fig. 13a,b.  (It should be noted that the routine for solving
Eqs. 11 does not converge well for large $\Delta_0$ near $x=0$, or for smaller
$\Delta_0$ at large x.)  There is a curious symmetry between the doping
of the vHs in the absence of spin-orbit coupling $x_{vHs}$ and that of the
diabolical point $x_{dia}$ in its presence: $x_{dia}\simeq -x_{vHs}$, Fig. 13c.
For the assumed value $t_{OO}^t=0.25eV$, the system is a semimetal at half
filling.  Figure 15 shows how the bands and Fermi level evolve with doping for
$\Delta_0^t=6eV$.  At $x=0.3$ (Fig. 15a), the Fermi level lies in the gap
between the two vHs at the $X$-point, but there is still a cylindrical Fermi
surface centered at $\bar M$, dashed line in Fig. 15d.  Reducing $x$ to 0.15
(Fig. 15b) or 0.05 (Fig. 15c), the Fermi level has now crossed one of the vHs
near $X$, leading to the presence of both electron and hole pockets, Fig. 15d.
\par
In Figure 16, the same sequence is repeated for $t_{OO}^t=0.1$.  As $x
\rightarrow 0^+$, $t_{OO}$ is still too large to have a zero-gap semiconductor,
in agreement with the scaling results, Fig. 12.  However, on the Cu side of the
charge transfer insulator ($x\le 0$), the much larger value of $\Delta$ leads
to
a smaller value of $y$, with a corresponding {\it zero-gap semiconductor at $x=
0^-$} -- i.e., the diabolical point falls at the Fermi level at (slightly less
than) half filling!  Figure 16e shows the band structure for $x=-0.05$ -- which
is as close to half filling as the numerical routine can be successfully used.
The diabolical point is clearly in a true gap in the dos.  The Fermi surface
for $x=-0.05$ is composed solely of an electron-like surface (dotted lines in
Fig. 16f) -- the lack of hole-like sections is further evidence that the net
electron concentration at half filling can vanish only by the Fermi surface
shrinking to a point (ZGS) and not by compensation of electron and hole
sections
(SM).

\subsection{Discussion: Mott Transition?}

\par
Early slave-boson calculations\cite{KLR}, which neglected direct O-O hopping
($t_{OO}=0$), identified the Mott transition as the point at which $t_{CuO}$ is
renormalized to zero.  Later calculations included direct oxygen-oxygen
hopping,
$t_{OO}$\cite{sad,RM3,Gri}, but it was found that slave boson calculations lead
to a metallic state at half filling\cite{New}, when the bare value $t_{OO}=
0.65eV$ is assumed.  As argued above, the value of $t_{OO}$ used in these
calculations should be a smaller, effective value, incorporating the role of
coupling to the $d_{z^2}$ states.  Moreover, the value of $N$ in the
relationship $t_{CuO}^{\prime}=\sqrt{N}t_{CuO}$ should, as argued above be
taken
equal to 1, and not 2, as is usually done.  When these two corrections are
made,
the present results give a metal-insulator transition at half filling, for
reasonable parameter values.
\par
This result is consistent with the results of other, non-slave boson
calculations of the three-band model. Thus, the
studies that produced the parameters of Table I\cite{Hyb,GraM,Esk,MAM} also
explored the low energy sector of the three-band model, and all find an
insulating state at half filling with a large charge transfer gap, comparable
to
experiment.  See also Refs. \cite{BC}, which find similar results.  For all of
these parameter sets, the slave boson calculation with $N=2$ would have
incorrectly predicted a metallic state at half filling.
\par
Zhang, et al.\cite{Eme} recently suggested that the slave boson calculations
are in error near half filling, and that the state at half filling is
always insulating (at least if $t_{OO}$ is not too large).  Furthermore, when
the slave boson calculation does predict an insulating state at half filling,
its detailed predictions are in good agreement with more accurate treatments.
It is still possible that, for finite $t_{OO}$, there is some minimum value
$\Delta_c$, below which
there is no metal insulator transition, since the Bi and Tl based cuprate
superconductors do not show evidence for an antiferromagnetic insulating phase.

\bigskip
\section{Conclusions}
\par
The present results have an important bearing on the question of the role of
the
vHs in driving structural transitions in the cuprate superconductors.  A
major stumbling block has been the assumption\cite{Poug} that the vHs
degeneracy
is not split in the LTO phase.  Spin-orbit coupling does produce such a
splitting, playing the role of the umklapp scattering postulated
earlier\cite{RM8A}.  However, vHs's cannot be destroyed in a two-dimensional
band, but only split up.  The ensuing Fermi surface reconstruction leads to a
large variety of new vHs's.  Perhaps the most interesting possibility is a
{\it diabolical point}, at which the Fermi surface collapses to a single point.
\par
While these Fermi surface reconstructions are undoubtedly present, just how
important they are depends on the proximity of the Fermi surfaces, as a
function
of doping, to the topological changes of the bands.  The answer to this
question, in turn, depends on the choice of the band parameters, and how they
are modified by correlation effects.  These are difficult questions, and the
present work is intended as an introduction to the problems involved, as well
as a {\it first estimate} for many of the parameters.
\par
Among the questions which depend on these parameter choices are the following.
Will the state at half filling be metallic or insulating?  Is the LTO
transition
driven by the vHs splitting?  At what doping does the vHs coincide with the
Fermi level.  Figure 11 provides an answer to the first question: the critical
value of $\Delta_0/t_{CuO}^{\prime}$ above which $t_{CuO}$ is renormalized to
zero by correlation effects, at precisely half filling.  The critical value
appears to be in the range of parameter estimates for both LSCO and YBCO.
The transition is {\it not} driven by
Fermi surface nesting, in that the transition always appears at half filling,
even though the nesting is better at other dopings.  However, there is an
element of nesting involved, in that the critical value of $\Delta_0$ increases
rapidly as $t_{OO}$ increases from zero.
\par
While I have demonstrated that the vHs degeneracy is split by the LTO
distortion, it remains to be seen whether the energy lowering associated with
this splitting can drive the structural transition.  This will be the case for
one special doping, when the vHs lies exactly at the Fermi level in the
undistorted structure.  However, I suggested\cite{RM8} a more interesting
possibility: that correlations can pin the Fermi level close
to the vHs over an extended doping range, so that the LTO transition could be
Fermi-surface-driven over this full range.  While this pinning is clearly
present, Fig. 13d, the Fermi surface at half filling is not close enough to
the vHs (Fig. 14) for this simple model to work.  However, the present
calculations suggest two closely related possibilities.
\par
One possibility is that a diabolical
point falls at half filling, so the half filled band is a zero-gap
semiconductor, and hence naturally an insulator at low temperatures.  If doping
then produces a simple rigid band filling, the resulting small hole pockets
can explain the observed Hall densities, and yet if the gaps are small the
Fermi
surfaces will appear to be large.  Sufficient hole doping would bring the holes
to a new vHs, associated with the box/coffin-like Fermi surface, Fig. 8b.
However, it seems that $t_{OO}$ is too large for the diabolical point to
fall at $x=0^+$.
\par
The second possibility, illustrated in Fig. 16, is that the diabolical point
falls at $x=0^-$.  In this case, the LTO structure would be stabilized both at
$x=0^-$ and at $x=x_{vHs}$, {\it but not at intermediate doping}.  In such a
situation, an electronic phase separation may arise, to keep part of the holes
at the vHs density\cite{RM3}.  This possibility arises for realistic values of
$t_{OO}$, but requires a somewhat enhanced value for the spin-orbit coupling
parameter $\lambda$.  However, in the
analogous one-dimensional metal problem, it is known that umklapp scattering is
strongly renormalized to drive the metal-insulator transition at half filling.
\par
A very attractive possibility is that {\it the renormalization arises from
antiferromagnetism}.  It is known\cite{LWH,Pick} that antiferromagnetism also
lifts the band degeneracy along the $X-M$ face of the Brillouin zone.
Attempts to self-consistently calculate antiferromagnetism -- to see if the
gap opening lowers the energy sufficiently to stabilize the antiferromagnetic
state -- generally find that the paramagnetic state is more stable.  Part of
the problem is that the gapping is imperfect: residual Fermi surface sections
are left behind.  If, because of spin orbit interaction, the system is already
close to a complete gap (a diabolical point), then it is possible that the
additional effect of antiferromagnetic correlations will open a complete gap,
and thereby enhance the stability of the antiferromagnetic state.  It should be
recalled that, upon doping away from the insulating state, the first carriers
produce small hole pockets near the $\bar M$ points, just as would be expected
for the diabolical point\cite{spinbag}.  In the problem of the quantum
Heisenberg antiferromagnet, it is known that Berry phases lead to dangerously
irrelevant couplings\cite{dang}.
\bigskip
\section{Discussion}
\par
Spin-orbit coupling introduces a major reconstruction of the Fermi surfaces of
the cuprates, and hence opens up a wide variety of questions, as well as
suggesting possible answers to a number of problems involving the cuprates.
Here, I briefly
indicate some of the major issues.  It should be noted that incorporation of
this scattering clearly demonstrates the important role of the vHs in the
structural transitions, in addition to their role in superconductivity.  In
brief, the LTO (and LTT) phase(s) are the equivalent of the charge-density wave
phases in the vHs problem, and the physics of the cuprates is dominated by the
competition between superconductivity and density-wave formation, in close
analogy to one-dimensional metals.

\bigskip
\subsection{Umklapp scattering and the Hall anomalies}
\par
It has previously been suggested that umklapp scattering could provide an
explanation for some of the anomalous features of the Hall effect in the
cuprates\cite{Hall}.  Basically, by introducing gaps at the vHs's, this
scattering would change the Fermi surfaces from large areas (proportional to
$1+x$) to much smaller areas, $\sim x/2$.  That spin-orbit coupling produces
just such an effect is clear from, e.g., Fig. 2, thus confirming the suggested
explanation.
\par
Moreover, the temperature dependence of the Hall density can, in principle, be
understood as a measure of the temperature-dependent gap associated with the
LTO phase transition.  As $T$ is lowered below the HTT-LTO phase transition,
the
gap opens gradually, converting the hole gas from the large-area Fermi surface
to the hole pockets, and then, as $T$ falls further, reducing the areas of the
pockets.

\bigskip
\subsection{Mott and Antiferromagnetic Transitions}
\par
This gradual gap opening can explain many anomalous features of the normal
phase above $T_c$, and particularly in the lightly doped regime near $x\simeq
0$.  This would be particularly true if, as a result of the combined effect of
spin-orbit coupling and large, Hubbard-like correlation effects, the Fermi
surface at half filling falls at a diabolical point, with vanishing area (as in
Fig. 4a,b).
\par
In particular, the mobility shows very little change at the LTO
transition, but there is a metal-insulator transition (change of sign of the
temperature coefficient of resistivity) at much lower temperatures.  This can
be
understood as another manifestation of a slow gap opening (enhanced in this
case
by the fact that the gap vanishes at the diabolical points).  Near the
transition temperature, the gap affects only the immediate vicinity of the
vHs's, but these holes make very little contribution to the mobility, because
of
the very strong inter-vHs scattering\cite{RMIV}.  Indeed, by eliminating the
vHs as a source of scattering, this could actually enhance the mobility of
holes on other parts of the Fermi surface.
\par
Gradually, however, the shrinking area of the residual Fermi surfaces with
decreasing $T$ would cause the mobility to decrease, ultimately leading to a
localization transition at half filling, when the areas shrink to zero.
\par
Just as in the usual localization picture, the spins should freeze out prior to
the final localization, leading to a local moment formation on the Cu's.  This
may explain why the Heisenberg model for the antiferromagnetic transition has
been so successful in these materials.  In the past, the success of this model
has led to suggestions for `two carrier dynamics' -- in which the holes on
the Cu were localized, while the doped holes on the oxygens formed a separate
band.  Here, it is seen that both features arise naturally from spin-orbit
scattering in the LTO phase.

\bigskip
\subsection{LTO-LTT Competition?}
\par
Since spin-orbit coupling directly splits the vHs degeneracy, it is not
necessary to postulate the existence of a dynamic JT effect, in order for the
LTO phase to be electronically driven.  Nevertheless, frozen phonon
calculations
find that the LTT distortion is energetically favored over the LTO or any other
local configuration\cite{Pick}.  Moreover, while spin-orbit scattering does not
strongly
affect the LTT-LTO splitting (Fig. 3a), the ordinary electron-phonon coupling
provides an additional energy lowering for the LTT phase (Fig. 3b).  Thus, the
possibility of dynamic JT scattering remains, and will be reanalyzed once the
dynamics of the transition to a uniform LTO phase is better understood.
\par
It must not be forgotten that even in molecular systems there is a strong
competition between the JT effect and spin-orbit coupling, and it will require
considerably more work to sort out their respective roles in the cuprates.

\bigskip
\subsection{Incommensurate Diffraction Peaks}
\par
Neutron diffraction studies have found that in La$_2$CuO$_4$, the
antiferromagnetic peaks are commensurate, at $Q=(\pi /a,\pi /a)$.  As the
material becomes Sr doped, the peaks split and become incommensurate, at
$Q\pm\delta (\pi /a,0)$ or $Q\pm\delta (0,\pi /a)$, with the incommensurability
$\delta$
increasing proportional to the doping $x$\cite{inco}.  A number of attempts
have
been made to interpret this as due to a peak in the electronic susceptibility
$\chi_q$ associated with a nesting feature in the Fermi surface, calculated
using the three band model, with $t_{OO}\ne 0$.
\par
There are certain difficulties associated with such interpretations, however.
In the first place, in a weak coupling calculation the intensity is orders
of magnitude larger for inter-vHs scattering (which in these models is
commensurate at $Q$) than the nesting-associated scattering when the Fermi
level is shifted off of the vHs (see Fig. 8 of \cite{BCT}).  In a marginal
Fermi
liquid model, this vHs intensity enhancement can be greatly reduced by
quasiparticle lifetime effects\cite{LZAM}.
In the doped material, the model predicts an incommensurate peak in the
susceptibility, in accord with experiment; however, the peak usually remains
incommensurate at half filling\cite{BeatL}.  To correct this problem, it is
necessary to assume\cite{LZAM} that the vHs's are very close to half filling
(since the nesting
feature that they produce is commensurate), and this is inconsistent with LDA
band structure calculations, if a semi-rigid band filling model is assumed.

\par
While the details of the doping dependence remain to be worked out, the present
model offers an alternative possibility for explaining the incommensurate
peaks: the vHs's are shifted off of the $X$ and $Y$ points, so that {\it
inter-vHs scattering is incommensurate}, and the peak positions depend on the
interplay of the correlation effects and tilting instability, which modify the
{\it shape} of the Fermi surface.
\par
Here, I can only indicate that the observed Fermi surfaces are consistent with
the direction and magnitude of the distortion $\delta$.  From Fig. 17a, it can
be seen that there are a large number (9) of inter-vHs scattering vectors near
$Q$, of which two are illustrated.  Figure 17b shows the relative
distribution of the nine peaks about the central commensurate position.  The
four outermost points match the positions of the observed incommensurate
neutron
scattering peaks (Fig. 17c).  As far as the magnitude of the effect goes, the
observation that $\delta\simeq 2x$\cite{inco} would require the vHs peak to
fall a fraction $x$ of the distance from, e.g., $Y$ toward $\Gamma$.  For $x=
0.14$, this is about twice as far as the largest Fermi surface of Fig. 8b.
\par
While this result is suggestive, a number of problems remain.
A more detailed calculation is required to determine the relative intensities
of
the various peaks.  Moreover, it is not clear how the doping dependence of the
peak arises in the present model.

\bigskip
\subsection{Phase Separation}
\par
An additional complication which has been neglected in this paper is the
possibility of some form of hole phase separation away from half filling, due
either to magnetic effects\cite{eme}, electron-phonon interaction\cite{RM3},
or long-range Coulomb interaction (specifically, the next-nearest-neighbor
Coulomb repulsion, $V$)\cite{Rai}.  In this transition, the holes
bunch up in such a way that part of them remain pinned at the insulating
phase at half filling.  The entries in Table II suggest yet another form of
instability, which may be called a `molecular JT' effect, since it arises when
$\Delta$ or $\Delta_d\rightarrow 0$, which leads to additional electronic
degeneracies.  There is considerable experimental evidence for such
phase separation\cite{EK,Recon,phassep}.  If a self-consistent theory for the
LTO structural transition can be developed, it will be possible to calculate
the criteria for many of these instabilities.
\par
The physics underlying the electron-phonon induced phase separation\cite{RM3}
can readily be understood.  The structural instability is stabilized by
lowering the kinetic energy of the {\it occupied} electronic states, and hence
the energy lowering is optimized when the Fermi level falls precisely at the
(undistorted) vHs -- or equivalently, half way between the two vHs of the
distorted structure.  In this case, the parameter regime associated with Fig.
16 is most interesting.  There are {\it two} dopings at which the Fermi level
falls at an optimal position for structural instability -- both at $x\simeq
0.15$, the original vHs, and at $x=0^-$, the diabolical point.  The new point,
$x=0^-$ will be associated with the greater structural instability (higher
transition temperature), since all of the dos is shifted away from the Fermi
level.  In contrast, for the old vHs, only part of the Fermi surface is gapped,
leaving behind hole pockets.  For intermediate dopings, the dos (in the
distorted phase) increases, leading to a reduced kinetic energy lowering.  If
this reduction is large enough, the system will prefer to phase separate,
keeping some holes at $x\simeq 0^-$ and the rest at $x\simeq 0.15$.
\bigskip
\subsection{Future Directions}
\par
Clearly, much work remains to be done, including working through detailed
calculations of the effects discussed in the earlier subsections of this
chapter.  It will also be important to study the temperature and doping
dependence of the LTO phase transition, in the presence of correlation effects.
Fitting the positions of the incommensurate neutron peaks may allow some of
the parameter values to be pinned down.  It would also be desirable to extend
the model to YBa$_2$Cu$_3$O$_{7-\delta}$, where such incommensurate modulations
have not been observed.

\par
{\bf Acknowledgments:}  I would like to thank G. Kotliar for useful
conversations.  Publication 583 from the Barnett Institute.

\bigskip
\par\noindent{\bf Appendix I: Band Dispersion Relations}
\bigskip\par
While the eigenvalue equation, Eq. 4, must be solved numerically, it is
convenient to rewrite it as a polynomial in $E$.  First, neglecting
electron-phonon coupling ($\eta =0$),
$$4E^2A_1+4ExA_2+x^2A_3+2(\Delta -E)B_1+(\Delta -E)^2B_2=0,\eqno(A1)$$
where $x=4t_{OO}$, $\bar s_i=sin(k_ia)$, and
$$A_1=\lambda_A^4-(|\lambda_x|^2\bar c_x+|\lambda_y|^2\bar c_y)^2+2t^2(2|
\lambda_x-\lambda_y|^2-\lambda_A^2)\bar c_x\bar c_y$$
$$+2t^2(\lambda_A^2+|\lambda_x
^2|\bar s_x^2+|\lambda_y|^2\bar s_y^2)+t^4(4-(\bar c_x+\bar c_y)^2),\eqno(A2)$$
$$A_2=\lambda_B^2(|\lambda_x|^2\bar s_x^2+|\lambda_y|^2\bar s_y^2)+t^2(|
\lambda_x+\lambda_y|^2+2t^2)(\bar s_x^2+\bar s_y^2)$$
$$-t^2(|\lambda_x|^2-|\lambda_y|^2)
(\bar c_x^2-\bar c_y^2),\eqno(A3)$$
$$A_3=\bar s_x^2\bar s_y^2(\lambda_B^2+2t^2)^2,\eqno(A4)$$
$$B_1=2E^3(\lambda_A^2+2t^2)+E^2x(\lambda_B^2+2t^2)(1+\bar c_x\bar c_y)$$
$$-E(x^2/2)
(|\lambda_x|^2\bar s_x^2+|\lambda_y|^2\bar s_y^2+t^2[\bar s_x^2+\bar s_y^2])
-(x^3/4)\bar s_x^2\bar s_y^2(\lambda_B^2+2t^2),\eqno(A5)$$
$$B_2=E^4-E^2(x^2/2)(1+\bar c_x\bar c_y)+x^4\bar s_x^2\bar s_y^2/16,\eqno(A6)$$
and $\lambda_A^2=|\lambda_x|^2+|\lambda_y|^2$,
$\lambda_B^2=\lambda_x\lambda_y^*
+\lambda_y\lambda_x^*$.
\par
Including electron-phonon coupling, the following terms should be added to
Eq. A1:
$$\eta^{\prime}C_1-\eta^{\prime 2}B_2,\eqno(A7)$$
where
$$C_1=4E^3(t^2[\bar c_x+\bar c_y]-|\lambda_x|^2\bar c_x-|\lambda_y|^2\bar c_y)
+4E^2xt^2(\bar c_x+\bar c_y)$$
$$+Ex^2(t^2[\bar c_x+\bar c_y][1-\bar c_x\bar c_y]
-\bar s_x^2\bar c_y|\lambda_x|^2-\bar s_y^2\bar c_x|\lambda_y|^2)-(x^3/8)
\lambda_B^2(\bar c_x+\bar c_y),\eqno(A8)$$
and $\eta^{\prime}=2\eta (\bar c_x+\bar c_y)$ in the LTO phase, or $\eta^
{\prime}=2\eta\bar c_x$ in the LTT phase, with $\eta_{LTT}=2\eta_{LTO}$.
It should be noted that, despite their complicated form, Eqs. A4,A7 are a
quadratic equation in $\bar c_y(E,\bar c_x)$, except in the LTO phase when
$\eta\ne 0$.
\par
These equations also have an important scaling form, in the narrow band limit.
That is, if $t$ is so small that $E>>E-\Delta$, then $E$ may be replaced by
the fixed value $\Delta$ everywhere in Eqs A4-A9, except when it occurs in the
combination $E-\Delta$.  In this case, scaling $t\rightarrow t^{\prime}=r_0t$,
$\lambda_i\rightarrow r_0\lambda_i$, for all $i$, $\eta\rightarrow r_0^2\eta$,
and $(E-\Delta)\rightarrow r_0^2(E-\Delta)$, the system of equations is {\it
invariant}.  That is, in the scaling limit, only the bandwidth is rescaled,
but the shapes of the Fermi surfaces are unchanged, and the dos keeps the
same shape, but is enhanced by a factor $1/r_0^2$.

\bigskip
\par\noindent{\bf Appendix II: Effective Value of $t_{OO}$}
\bigskip\par
\par\noindent{\bf IIa: Equation for $t_{OO}^{eff}$}
\bigskip\par
Feiner, et al.\cite{FGD} introduced a five band model of the CuO$_2$ plane, to
explore the role of the apical oxygen.  Here, I use the corresponding
one-electron band structure to demonstrate the reduction in value of the
effective O-O hopping parameter, $t_{OO}$.  The band dispersion can be
written in the form
$$A(1-b[s_x^2+s_y^2]-cs_x^2s_y^2)=0,\eqno(B1a)$$
with
$$A=E^2(E-\Delta )V,\eqno(B1b)$$
$$Ab=E[(E-\Delta )W+4Et_{CuO}^2V],\eqno(B1c)$$
$$Ac=8[t_{OO}(E-\Delta )+2t_{CuO}^2][2t_{OO}V-W],\eqno(B1d)$$
$$V=(E-\bar\Delta_z)(E-\delta_O)-\alpha_1^2,$$
$$W=4[(E-\bar\Delta_z)\alpha_2^2+(E-\delta_O)\alpha_3^2-
2\alpha_1\alpha_2\alpha_3].$$
The definition of the parameters differs somewhat from Feiner, et
al.\cite{FGD}.
The zero of energy is taken at the planar O $p$-levels, and $t_{OO}$ is taken
as
a positive number.  The new parameters are $\bar\Delta_z=\Delta -\Delta_z$, the
Cu $d_{z^2}$ level, $\delta_O$, the apical O level, $t_{CuA}=\sqrt{3}\alpha_1$,
the Cu $d_{z^2}$ -- apical O hopping parameter, $t_{OA}=\alpha_2/2$, the
apical O -- planar O hopping parameter, and $t_z=\alpha_3$, the Cu $d_{z^2}$ --
planar O hopping parameter.  Molecular orbital theory would give $t_z=t_{CuO}/
\sqrt{3}$.  Calculated\cite{GraM} values for these parameters are listed in
Table I.
\par
{}From Eq. B1a, the vHs is determined as that energy for which the point (e.g.)
$s_x=1$, $s_y=0$ is at the Fermi level, or $b=1$.  For this choice of $b$,
the shape of the Fermi surface, $k_y(k_x)$ is given by the solution of Eq. B1a.
Hence, {\it all models with the same value of $c$ will have the same shape
of Fermi surface at the vHs}!  In particular, the vHs will fall at exactly the
same doping.  Thus, for a three-band model to agree with the five-band model,
the parameter $t_{OO}$ must take different values in the two models.  [More
generally, the three-band $t_{OO}$ will have a different value from the LDA
$t_{OO}$ -- presumably, the five-band model will be closer to the LDA value.]
\par
In the three-band model, the band dispersion is given by Eq. B1a, with
$$c=2y+y^2,$$
and $y=4t_{OO}^{eff}/E_{vHs}$.  Hence the vHs of the three-band model will fall
at the same doping as that of the five-band model, if $t_{OO}^{eff}$ is chosen
to satisfy
$$y=y^{\prime}-{2W\over E_{vHs}^{\prime}V},\eqno(B2)$$
where $y^{\prime}=4t_{OO}/E_{vHs}^{\prime}$ in the five-band model.  Note that
not only is $t_{OO}\ne t_{OO}^{eff}$, but the vHs is located at a different
energy.  In the five-band model, the vHs is given by the largest zero of $b=1$,
or
$$(E[E-\Delta ]-4t_{CuO}^2)V-(E-\Delta )W=0,\eqno(B3)$$
while in the three-band model, the vHs is given by Eq. B3 with $W=0$, $V=1$.
Defining $\eta_{vHs}=E_{vHs}/E_{vHs}^{\prime} (\le 1)$, then Eq. B2 can be put
into a suggestive form when $\alpha_1=0$:
$$t_{OO}^{eff}=\eta_{vHs}\bigl(t_{OO}-{2\alpha_2^2\over(E-\delta_O)}-{2\alpha_3^2
\over (E-\bar\Delta_z)}\bigr).\eqno(B4)$$
This is essentially the form suggested in Table II.
\par
Note that the above reasoning implicitly assumes that the only role of $t_{OO}$
in the theory is to move the vHs away from half filling.  This is in fact my
present belief.  If it is found that $t_{OO}$ plays some other role (or that,
e.g., the apical oxygen plays such a role), then this would be evidence that
the simple three-band model cannot capture the essential physics of the
process.  Until such time as that becomes necessary, however, it is simpler to
work with the three-band model, with correctly chosen effective parameters.
\bigskip\par
\par\noindent{\bf IIb: Numerical Estimates of $t_{OO}^{eff}$}
\bigskip\par
Given values for all the band parameters, $t_{OO}^{eff}$ can be calculated from
Eq. B2.  Once more, there
are separate problems in estimating the bare parameters and the renormalized
parameters.  Here, I will only give an illustration of the orders of magnitude
expected.  The new parameters will be assumed to have the values given in Table
I\cite{GraM}, with $\Delta =4eV$, $\bar\Delta_z=3.36eV$, $t_{CuO}=1.3eV$, and
$t_{OO}=0.65eV$.  Since the effect of each of the three hopping terms is quite
different, I first illustrate how $t_{OO}^{eff}$ would change if only one of
the
$\alpha$'s were non-zero (with value given by Table I).  If only $\alpha_1$ is
non-zero, it has a small effect on the vHs, $t_{OO}^{eff}=0.57eV$.  Either
$\alpha_2$ or $\alpha_3$ have comparable effects, with either leading to
$t_{OO}^{eff}=0.40eV$.  For all three $\alpha$'s non-zero, $t_{OO}^{eff}=0.28
eV$.  Hence, the expected reductions can be quite substantial.  For somewhat
larger $\alpha$ values, $t_{OO}$ could even change sign.  Thus, for $\alpha_2=
1$, $\alpha_3=0.75$ (the molecular orbital value), and $\alpha_1=0$,
$t_{OO}=-0.1$.
\par
Correlation effects will modify this result, particularly near half filling.
In particular, both $\alpha_1$ and $\alpha_3$ should scale to zero as $\sim r_
0$, but $\alpha_2$, which involves an O-O hopping, should not be much affected.
Since $\Delta$ is itself reduced by a factor of $\sim 3$, and it enters into
the denominator of Eq. B2 or B4 ($E\simeq\Delta$), then $t_{OO}^{eff}$ may not
be significantly renormalized.
\par
In principle, going from a five-band to a three-band model could also
renormalize the total width of the band -- i.e., lead to an effective value for
$t_{CuO}$ as well.  In practice, this does not seem to occur, so only $t_{OO}$
is renormalized.
\bigskip
\par\noindent{\bf Appendix III: Distortion of CuO$_6$ Octahedra}
\bigskip\par
Ohta, et al.\cite{Ohta}, by compiling data on a large number of cuprate
superconductors, were able to find only one consistent correlation between
$T_c$ and a molecular property: the larger the apical oxygen -- planar oxygen
energy splitting, $\Delta V_A$, the higher is $T_c$.
In the text, I suggest a related correlation: a smaller value of the $d-d$
splitting $\Delta V_d$ should cause a smaller effective value of $t_{OO}
$\cite{AAA}, which in turn correlates with a lower value of $T_c$.
\par
The correlation of $\Delta V_d$ with $d_a$ has an origin reminiscent of a
`classical' JT effect, and indeed Cu is a JT ion, often found in a distorted
octahedral environment\cite{JT}.  That is, if the six oxygens around each
Cu formed a perfect octahedron, then the Cu $d_{x^2-y^2}$ and $d_{z^2}$ levels
would be exactly degenerate.  The structural distortion lifts the degeneracy,
as
in a static JT effect.  However, in the ordinary JT effect it is the splitting
of the electronic degeneracy which is the driving force for the distortion.
Here the problem is more complex -- there are large commensurability strains
between the Cu planes and the La planes, and the octahedral distortion leads to
a considerable strain relief (this is why all octahedra distort along the same
direction).  The same sort of distorted octahedra appear in LaNiO$_4$, and
Ni is not a JT ion\cite{Pick}.
\par
Nevertheless, the distortion does cause the d-d splitting, as can be seen from
an analysis of the data base of Ohta, et al.  In particular, $\Delta V_d$ is
linearly proportional to $d_A/d_p$, the octahedral distortion ($d_p$ is the
planar Cu-O distance), Figure 18a.
Most of the compounds analyzed have only a five-fold oxygen coordination about
the Cu, so even when $d_A/d_p=1$, the two $d$ levels are not degenerate.  That
is why the solid line in Fig. 18a does not pass through $\Delta V_d=0$ when
$d_A=d_p$.  For the 6-fold coordinated materials (circled in Fig. 18a), it is
possible to draw an alternative line (dashed line) which does pass through this
point.
\par
Now, when $d_A\ne d_p$, the apical and planar O's will have different Madelung
energies, so there should be a correlation between $\Delta V_A$ and $d_A/d_p$.
Unfortunately, this correlation is considerably weaker (Fig. 18b) than that
with $\Delta V_d$, presumably because the apical O's see the
non-CuO$_2$ planes, which vary considerably from one material to another.
\par

\bigskip
\par\noindent{\bf Appendix IV: Electron Picture of Correlation Effects}
\bigskip\par
Equations 11, which are valid in a hole picture, can be replaced by the
following equations, in an electron picture:
$$r_0^2={1\over 2}[\sum_ku_k^2f_e(E_k)-1],\eqno(D1a)$$
$$\Delta_0-\Delta ={1\over 2r_0^2}\sum_ku_k^2f_e(E_k)(\Delta-E_k).\eqno(D1b)$$
The equivalence of the two pictures can be shown by subtracting the
corresponding pairs of equations (e.g., Eq. 11a and D1a) and using the fact
that
$f_e(E_k)=1-f_h(E_k)$ (valid for all temperatures).  This results in
$${1\over 2}\sum_ku_k^2=1,\eqno(D2a)$$
$$\Delta ={1\over 2}\sum_ku_k^2E_k.\eqno(D2b)$$
\par
In the above equations, it is essential to keep in mind that the summation is
over all $k$-states of {\it all three hybridized bands}, Eq. 12.  Equations D2
have a simple interpretation, saying that the three bands contain exactly one
Cu
electron (and two O electrons) per unit cell (Eq. D2a), and that the average Cu
energy is unchanged by hybridization (Eq. D2b).  To prove Eq. D2b, the explicit
form of $u_k$ and $E_k$ are required.  For simplicity, only the simplest case,
based on Eq. 1 with $t_{OO}$ (and $U$) set equal to zero.  In this case, the
nonbonding band is purely O-like, and may be neglected.  For the bonding and
antibonding bands,
$$E_k={\Delta\over 2}\pm\sqrt{({\Delta\over 2})^2+4t_{CuO}^2\gamma_k},
\eqno(D3a)$$
$\gamma_k=sin^2(k_xa/2)+sin^2(k_ya/2)$, and
$$u_k^2={E_k\over 2E_k-\Delta}.\eqno(D3b)$$
Using Eqs. D3,
$$\sum_ku_k^2E_k=\sum_k{}^{\prime}{E_k^2\over 2E_k-\Delta}$$
$$={1\over 2}\sum_k{}^{\prime}[E_k+{\Delta\over 2}(1+{\Delta /2\over E_k-\Delta
/2})]$$
$$=[\Delta +{\Delta\over 2}(2+0)]=2\Delta ,$$
thus confirming Eq. D2b.
Here the prime means summation over two bands only.
\par
In the above argument, it is essential to note that the equivalence works only
if the summation is over all three bands (in the more general case $t_{OO}\ne
0$).  Hence, at $T\simeq 0$ the hole picture is considerably more convenient,
since the summation can in this case be restricted to the antibonding band
only.

\vskip 0.3in
\begin{tabular}{||c||r|r|r||}        \hline
\multicolumn{4}{c}{{\bf Table I: Bare Parameters (eV)}} \\ \hline\hline
\multicolumn{1}{c}{Parameter} &
  \multicolumn{3}{c}{Reference} \\   \hline
\multicolumn{1}{c}{ } & \multicolumn{1}{r}{\cite{Hyb}} &
  \multicolumn{1}{r}{\cite{GraM}} & \multicolumn{1}{r}{\cite{Esk}} \\
    \hline\hline
$\Delta$ & 3.6 & 3.51 & 3.5     \\     \hline
$t_{CuO}$ & 1.3 &  1.47 & 1.3   \\     \hline
$t_{OO}$ & 0.65 &  0.61 & 0.65   \\     \hline
$V$ & 1.2 & 0.52   & $<1$       \\     \hline
$\Delta_d$ & -- &  0.64 & --     \\     \hline
$\delta_0$ & -- &  1.46 & --     \\     \hline
$\alpha_1$ & -- &  0.47 & --     \\     \hline
$\alpha_2$ & -- &  0.66 & --     \\     \hline
$\alpha_3$ & -- &  0.50 & --     \\     \hline
\end{tabular}
\vskip 0.3in
\begin{tabular}{||c||c|c|c||}        \hline
\multicolumn{4}{c}{{\bf Table II: Suggested Effective Parameter Values}} \\
    \hline\hline
{Parameter} &Bare Value & Scaling$^*$ & Corrected Bare Value
     \\   
  & (eV) & & (eV) \\
    \hline\hline
$t_{CuO}$ & 1.3 & $\sim\sqrt{N^{\prime}} r_0$ (Eq. 11a) & \\     \hline
$\Delta$ & $\Delta_0+2V$ & (Eq. 11b) & 5-6  \\   \hline
$t_{OO}$ & $t_{OO}^1\simeq 0.65$ & $t_{OO}^1-X/\Delta$  & 0.14 (LSCO)
    \\     
  &  &  & 0.3  (YBCO)    \\     \hline
$\eta$ & 0.005 (LTO) &  $\sim N^{\prime}r_0^2\theta^2/\Delta$ &
0.016 (LTO) \\
       & 0.08  (LTT) &                      &  0.25  (LTT) \\     \hline
$\lambda$ & 0.006 (LTO) & $\sim\sqrt{N^{\prime}} r_0\theta /\Delta_d$ &
? \\     
          & 0.024 (LTT) &                     &           \\     \hline
\noalign{*: $N^{\prime}=N/2$ in the spin-corrected model.}
\end{tabular}

\bigskip
\centerline{\bf Figure Captions}
\bigskip
{\bf Fig.~1} Energy dispersions of the LTO phase of the 2D CuO$_2$ sheets, for
various choices of band parameters.  [Dot-dashed lines = dispersion along the
line $X\rightarrow\bar M$.]
Parameters are $t=1.3eV$, $\Delta =4eV$, $t_{OO}=0.65eV$ in
all cases, with $\lambda_2$, $\lambda_1$, $\eta$ = (a) 0, 0, 0; (b) 6meV, 0, 0;
(c) solid lines = 50meV, 0, 0; dashed lines = 50meV, 0, 100meV;
(d) 50meV, 50meV, 0.

\smallskip
{\bf Fig.~2} Fermi surfaces of the LTO phase of the 2D CuO$_2$ sheets, assuming
the same parameters
as Fig. 1a,b, respectively.  The Fermi energies are (following the solid
surfaces of Fig. 2a, from lower left to upper right) $E_F$ = 4.5, 5.0, 5.28,
and 5.5eV.

\smallskip
{\bf Fig.~3} Comparing the Fermi surfaces of the LTT (solid and dashed lines)
and the LTO phases (dotted lines) [dashed lines = dispersion near $Y$; solid
lines = near $X$]. (a) = same parameters as Fig. 1c (solid lines); (b) = same
parameters as Fig. 1c (dashed lines).

\smallskip
{\bf Fig.~4} Energy dispersion in the LTO phase, for several values of
$t_{OO}$.
Parameters are the same as the solid lines in Fig. 1c, except $t_{OO}$ = (a) 0,
(b) 0.1, (c) 0.25, and (d) 0.65eV [N.B.: Fig. 4d is identical to Fig. 1c, solid
lines].

\smallskip
{\bf Fig.~5} Density of states (plotted against electron density) for the same
four $t_{OO}$ values as in Fig. 4.

\smallskip
{\bf Fig.~6} Fermi surfaces associated with changes of topology (vHs or
diabolical point), for the same parameters as Figs. 4,5.  (a) $E_F$ =
5.225 or 5.337eV (solid line -- the hole-like and electron-like Fermi surfaces
have identical shape), $E_{dia}=5.282eV$; (b) 5.226 (dotted line), 5.338
(dashed), and 5.34eV (solid), $E_{dia}=5.322eV$; (c) 5.227 (dotted), 5.337
(dashed), 5.386 (dot-dashed), and 5.393eV (solid); (d) 5.230 (dotted), 5.337
(dashed), 5.581 (dot-dashed), and 5.583eV (solid).  The dot-dashed lines are
the
diabolical points.

\smallskip
{\bf Fig.~7} Carrier density $n$ (electrons per Cu) for the same parameters as
in Figs. 4-6:  from left to right, $t_{OO}$ = 0, 0.1, 0.25, and 0.65eV.

\smallskip
{\bf Fig.~8} Evolution of Fermi surfaces near primary hole vHs. (a) Parameters
of Fig. 6d ($t_{OO}=0.65eV$), for several energy values, $E_F$ = 5.21 (dotted
line), 5.228 (dot-dash), 5.23 (solid), and 5.25eV (dashed).
(b) vHs for the four cases of Fig. 6: $t_{OO}$ = 0 (solid line), 0.1 (dashed),
0.25 (dot-dashed), and 0.65eV (dotted).

\smallskip
{\bf Fig.~9} Fermi surfaces associated with changes of topology in the LTT
phase, for the same parameters as Fig. 3b.  Energies are $E_F$ = 4.96 (dashed
lines), 5.35 (dotted), 5.47 (solid), and 5.58eV (dot-dashed).
\smallskip

\smallskip
{\bf Fig.~10} Doping $x_{vHs}$ at which the Fermi level coincides with the vHs,
vs $t_{OO}$, assuming $t_{CuO}=1.3eV$.  Bare parameters = solid ($\Delta =4eV$)
and dashed ($\Delta =6eV$) lines; renormalized parameters = open ($\Delta_0=4
\sqrt{2}eV$) and filled ($\Delta_0=6\sqrt{2}eV$) squares; alternatively
normalized
parameters ($t_{OO}^t$) = open ($\Delta_0^t=4eV$) and filled ($\Delta_0^t=6eV$)
circles.  Dashed lines = guides to the eye.  Solid horizontal lines = expected
values of $x_{vHs}$ for LSCO and YBCO.

\smallskip
{\bf Fig.~11} Threshold of metal-insulator transition.  Open circles = full
slave boson calculation; lines = scaling regime.  Parameters ($\lambda_2,\eta_0
$) = (0,0) (solid line and open circles); (0.005eV,0) (dashed line); (0.05eV,0)
(dot-dashed line); (0.05eV,0.1eV) (dotted line).

\smallskip
{\bf Fig.~12} Semimetal (SM) zero-gap semiconductor (ZGS) crossover, in scaling
regime, assuming $\lambda_1$ = 0 (solid line) or $\lambda_1$ =$\lambda_2$
(dashed line). In the scaling regime, $y=8t_{OO}/\Delta_0$.

\smallskip
{\bf Fig.~13} Slave boson calculation of metal-insulator transition, assuming
$\Delta_0^t$ = 4 (solid line), 5 (dashed line) or 6eV (dot-dashed line):
(a) $t_R$; (b) $E_F$; (c) $x_{vHs}$ (for the lines) or $x_{dia}$ (for the
symbols); (d) $\Delta E$ = $E_F-E_{vHs}$ (lines) or $E_F-E_{dia}$ (symbols).
The dotted line in Fig. 13a is the scaling regime approximation.  The vHs is at
the Fermi level when the corresponding line crosses the dotted line in Fig.
13c.

\smallskip
{\bf Fig.~14} Fermi surfaces corresponding to $\Delta_0^t=6eV$ in Fig. 13
(dot-dashed lines), for $x$ = $x_{vHs}$ (dashed line), $0^+$ (solid line), or
$0^-$ (dot-dashed line).

\smallskip
{\bf Fig.~15} Renormalized band structure and Fermi surfaces corresponding
to $\Delta_0^t=6eV$, $t_{OO}^t=0.25eV$, $\lambda_2^t=0.05eV$, for various
dopings $x$ = 0.3 (a), 0.15 (b), or 0.05 (c).  Horizontal line = Fermi level.
Fig. 15d = corresponding Fermi
surfaces: $x$ = 0.3 (dashed line), 0.15 (dot-dashed), 0.05 (solid).

\smallskip
{\bf Fig.~16} Renormalized band structure and Fermi surfaces corresponding
to $\Delta_0^t=6eV$, $t_{OO}^t=0.1eV$, $\lambda_2^t=0.05eV$, for various
dopings $x$ = 0.25 (a), 0.15 (b), 0.05 (c), 0.02 (d), r -0.05 (e).  Solid
horizontal line = Fermi level; dashed line in Fig. 16e = diabolical point.
Fig. 16f = corresponding Fermi
surfaces: $x$ = 0.25 (dashed line), 0.15 (dot-dashed), 0.05 (solid), -0.05
(dotted).

\smallskip
{\bf Fig.~17} (a) Schematic of full Brillouin zone, corresponding to Fig. 8b,
illustrating some possible incommensurate, inter-vHs scattering vectors. (b)
Corresponding distribution of all possible scattering vectors around central
commensurate peak $Q=(\pi /a,\pi /a)$. (c) Observed incommensurate diffraction
pattern in LSCO\cite{inco}; $+$ indicates position of $Q$.

 \smallskip
{\bf Fig.~18} Energy level splittings, either Cu $d_{z^2}-d_{x^2-y^2}$ (a) or
apical vs. planar O (b), plotted against normalized Cu-apical O distance,
$d_A/d_p$.  Letters refer to different superconducting compounds, after Ohta,
et
al.\cite{Ohta}.  Circled letters are compounds with 6-fold (distorted
octahedral) coordination of the Cu.

\end{document}